\documentclass[12pt]{article}

\usepackage{hyperref}
\usepackage{epsfig}
\usepackage{amsmath}
\usepackage{amsfonts}
\usepackage{amssymb}
\usepackage{eufrak}
\usepackage{mathrsfs}

\def\arXiv#1{\href{http://xxx.lanl.gov/hep-th/abs/#1}{#1}}

\input epsf.sty
\renewcommand{\theequation}{\thesection.\arabic{equation}}

\def\be{\begin{equation}}
\def\ee{\end{equation}}
\def\ba{\begin{eqnarray}}
\def\ea{\end{eqnarray}}
\def\mc{\mathcal}

\def\appendix{{\section*{Appendix}}\let\appendix\section%
        {\setcounter{section}{0}
        \gdef\thesection{\Alph{section}}}\section}

\begin{document}

\thispagestyle{empty}
\def\thefootnote{\fnsymbol{footnote}}
\begin{flushright}
 RUNHETC-2003-29 \\
 hep-th/0310106 \\
 \end{flushright}
\vskip 0.5cm

\begin{center}\LARGE
{\bf Renormalization Group Approach to $c=1$ Matrix Model on a
circle and D-brane Decay}\\

\end{center}

\vskip 1.0cm

\begin{center}
{\centerline{Satabhisa Dasgupta\footnote{{\tt
satavisa@physics.rutgers.edu}}}}

\vskip 0.5cm

{\it Department of Physics and Astronomy, Rutgers University,
\\ Piscataway, NJ 08854, U.S.A.}
\end{center}

\vskip 0.5cm

\begin{center}
{\centerline{Tathagata Dasgupta\footnote{{\tt
dasgupta@physics.nyu.edu}}}}

\vskip 0.5cm

{\it Department of Physics, New York University,
\\ 4 Washington Place, New York, NY 10003, U.S.A.}
\end{center}

\vskip 0.75cm


\begin{abstract}

Motivated by the renormalization group (RG) approach to $c=0$
matrix model of Bre\'zin and Zinn-Justin, we develop a RG scheme
for $c=1$ matrix model on a circle and analyze how the two
coupling constants in double scaling limit with critical exponent
flow with the change in length scale.  The RG flow equations
produce a non-trivial fixed point with the correct string
susceptibility exponent and the expected logarithmic scaling
violation of the $c=1$ theory. The change of world-sheet free
energy with length scale indicates a sign change as we increase
the temperature, indicating a phase transition due to liberation
of the non-singlet states. At low temperature, the RG analysis
also lead to T-duality of the singlet sector free energy. The RG
flow to the $c=1$ fixed point can be understood as the decay of
unstable $D0$-branes with open string rolling tachyon to the $2D$
closed string theory described by the end point of the flow. The
amplitude of the decay is extracted from the change of the
world-sheet free energy described by the RG process and is in
accordance with the prediction from boundary Liouville theory.

\end{abstract}

\vfill \setcounter{footnote}{0}
\def\thefootnote{\arabic{footnote}}
\newpage



\tableofcontents

\section{Introduction}
\setcounter{equation}{0}

Back in the late 80's and early 90's, there had been extensive
work on low dimensional string theories, namely the $c\le 1$
non-critical (or $D \le 2$ critical) bosonic string theories, both
in the discretized large $N$ matrix model
approach~\cite{VK,KKM,ADF,FD2} and in continuum Liouville theory
approach~\cite{KPZ,FD1, DK} by looking at them as $2D$ quantum
gravity coupled to $c \le 1$ matter. One of the motivations was to
learn something about the non-perturbative effects of $2D$ quantum
gravity and string theory and to apply that knowledge to higher
dimensions. Studying $2D$ gravity with simple matter was thought
to be useful to probe issues of topology change in a finite and
tame gravitational theory and to construct a string theoretical
basis of QCD. A remarkable fact is that all the ideas of
developing non-perturbative techniques using branes and dualities,
string field theory {\it etc.} is now going back to matrix model.
According to the recent idea, initiated in \cite{mgv}, the
eigenvalues of $c=1$ matrix model represent unstable $D$-branes in
the dual two-dimensional string theory. Before going into the
details of this idea in the context of this paper, we describe
briefly some of the difficulties in understanding $c=1$ matrix
model on compact space to sketch the main motivation of this
paper.

In the discrete approach (see
\cite{kazakov,IK-rev,ginsparg-moore,FGZ-2d,jevicki-rev,Polchinski-rev}
for review), the Euclidean path-integral
for two dimensional gravity can be written
as a sum over discretized random surfaces and is evaluated using
the large $N$ or the planar limit of the appropriate random matrix
models. However the large $N$ limit, the lowest order in the
string perturbation theory, is not the only thing that we can
extract from the expansion of the world-sheet free energy.
In the continuum limit,
for any genus, the area of the surface is large compared to
the elementary polygons of the discretized surface. Hence for any
fixed genus $G$, {\it i.e.}, at order $N^{-2G}$ in the
$1/N$-expansion, we should stay in the vicinity of the singular
point $g = g_c$ of the free energy at which the mean area
diverges. Thus the limit of interest involves large $N$ and small
$(g-g_c)$ with $g_s^2
\equiv \frac {1} {N^2 (g-g_c)^{2-\gamma_{str}}}$ (where
$\gamma_{str}$ is the string susceptibility exponent) fixed.
This {\it double scaling limit}
\cite{brezin-kazakov,douglas-shenker,gross-migdal} led to
the solution of the matrix models to all orders in string
perturbation theory. The genus expansion of the world-sheet
free energy can be written as

\be F(N,g)=\sum_{G=0}^{\infty}
N^{2-2G}(g-g_c)^{(1-G)(2-\gamma_{str})} \,. \ee Also the genus
expansion becomes universal, {\it i.e.} almost entirely
independent of how the surface is discretized.

The simplest model involving integral over one matrix
variable has been solved to describe pure two-dimensional gravity
(the $c=0$ case)
\cite{brezin-kazakov,douglas-shenker,gross-migdal}. The
one-dimensional hermitian matrix chain models have been solved in
the double scaling limit and identified with the $c<1$ minimal
models coupled to two-dimensional gravity
\cite{bdks,gm-PRL,cgm,douglas-KdV}. For this general case of
$(p,q)$ minimal models coupled gravity, Douglas has proposed a
solution in terms of the generalized KdV equations
\cite{douglas-KdV}, where the non-perturbative partition function
is given by the square of the $\tau$-function of the KdV
hierarchy, satisfying the string equation.
The hermitian matrix quantum mechanics,
describing $c=1$ theory, has been solved
\cite{GrossMiljkovic,BKZamo,GinsZ-J,Parisi} and is
interpreted in terms of two-dimensional string theory
\cite{DasJevicki,Polchinski-collective}, where the role of the
extra dimension is played by the conformal factor of the
world-sheet quantum gravity.

The solvability of the $c\le 1$ matrix models of $N^2$ degrees of
freedom is mainly due to the fact that only $N$ eigenvalues
contribute, and the other $N(N-1)$ angular degrees of freedom
decouple. The $SU(N)$ singlet spectrum is described by $N$ free
fermions moving in a potential. The non-singlet
states, coming from the angular degrees of freedom have energy
diverging as logarithm of the cut-off. So they are pushed to
infinity in the continuum limit. As a result, the singlet wave
functions of the $c=1$ matrix model are sufficient to account for
all physical states of the continuum string theory in a
non-compact dimension, or in a circle bigger than a critical size,
$R>R_c$ \cite{GK1}. This decoupling corresponds to the confinement
of Berezinskii-Kosterlitz-Thouless (BKT) vortices
\cite{GK2,BoulKaza}, which wind around the target space plaquette
of lattice size. Although each of the vortices is suppressed by a
divergent action in the continuum limit, its entropy factor, given
by the number of places it can be found, grows and for
sufficiently small circle $R<R_c$, a Kosterlitz-Thouless phase
transition occurs. As a result, although exact results in matrix
model have been proved to be very powerful in describing low
dimensional string theory in uncompactified dimension,
understanding physics from very small compact target space
is lacking. Although, for $c=1$ case, $X$ and the extra hidden
dimension or the Liouville field $\phi$ are {\it a priori} spatial
dimensions, it is more convenient to continue $X\to iX$, and
denoting $t = iX$ as the Euclidean time. If we have $t$ wrapped on a
circle, $t \sim t+2\pi nR$, we have $c=1$ string at finite
temperature.

For $c>1$, all the $N^2$ angular degrees of freedom become
relevant and the situation becomes difficult to pursue in the
matrix model framework, until we understand considerably how to
deal with the non-singlet states. At this moment the predictions
of continuum theories for the well-understood $c\le 1$ cases turn
out to be meaningless for $c>1$, as the KPZ-DDK formula
\cite{KPZ,FD1,DK} for the string susceptibility exponent and for
scaling dimensions of matter operators lead to nonphysical complex
values. Having said all these, the difficulties with $c>1$ models
are not related to just these technical issues. The continuum
approach indicates that these models are tachyonic
\cite{seiberg,KutSei}. Light-cone quantization of certain $c=2$
matrix model gave some partial results \cite{DalleyKlebanov}, but
at the same time faces potential difficulties as the non-singlet
states do not confine even if non-compact target space is
considered.

The motivation of this paper is to initiate an approach to deal with
the less understood non-singlet sector in matrix quantum
mechanics on a compact space.
We develop a renormalization group (RG) scheme,
generalizing the work of Br\'ezin and
Zinn-Justin~\cite{brezin-zinnjustin}  on $c=0$
to the $c=1$ model on a circle.  The
motivation is to develop a scheme which would reasonably reproduce
the known cases that are solvable and to understand, at least
qualitatively, the physically interesting situations which can not
be simplified because of the lack of the solvable structure.
The RG analysis of the $c=0$ model, the simplest of the
solvable examples, studied in \cite{brezin-zinnjustin}
has been extended to various general cases
(see for example \cite{brezinhikami,itoi,fdavid,bonnet-david}).
In \cite{itoi}, reparametrization invariance and the the loop equations are
used
to eliminate some of the induced interactions which appear in the
RG transformation at higher orders in coupling. However extension of this
method
to matrix quantum mechanics on circle is not obvious.
We instead directly  generalize \cite{brezin-zinnjustin} by explicitly
evaluating the determinant obtained by integrating out part of the
matrices to get a RG flow in the space of actions in
general representation. This enables us to study the models which do not
allow
angular integration, such as the $c=1$ model on small circle, or $c > 1$
matter coupled to gravity.

As an illuminating interpretation and feedback from our RG
analysis, we will digress to the recent progress in understanding
the dynamics of the boundaries in the $2D$ string theory, that
realizes the duality between $c=1$ model and $2D$ quantum gravity
coupled to $c=1$ matter as an exact open/closed string duality
\cite{mgv,martinec,kms,mgtv,schomerus,akk,sen-kyoto03}. According
to this idea, the quantum mechanics of $SU(N)$ invariant matrix
variables in an inverted oscillator potential is visualized as the
quantum mechanics of open string tachyons attached to $N$ unstable
$D0$ branes that decays into ({\it i.e.} dual to) Liouville theory
coupled to $c=1$ matter describing $2D$ closed string theory
together with its $D0$ branes. In matrix model, an unstable brane
corresponds to a free fermion excited to the top of the potential
(in presence of the rest forming the static Fermi sea).  It then
decays to the closed string vacuum by rolling down to the Fermi
level as an unstable eigenvalue trajectory
$$
 z(t)=\sqrt{2 \mu \alpha'}~\hat
\lambda~ e^t \,.
$$
The amplitude of this decay calculated
by continuum methods can be read from the matrix model
analysis considering the bosonization of the relativistic
fermions in the asymptotic limit  \cite{kms}.
Alternatively as proposed in \cite{mgtv}, one can
formally view the Vandermonde determinant
corresponding to the
$(N+1)$-th unstable eigenvalues in presence of the rest $N$,
to be related to the bosonization field and thus to lead to
the decay amplitude.

To understand from our
RG analysis, we consider integrating out one of the
eigenvalues in presence of the static Fermi sea
of the rest of the $N$ eigenvalues. The decay amplitude is shown
to be contained in the change of the world-sheet free energy given by the
ratio
$Z_{N+1}/Z_{N}$ of the partition functions. The nice thing is that the
end point of such a decay is explicitly a $c=1$ fixed point of the flow.
We compute the above-mentioned Vandermonde
determinant, the entropy arising out of
the loss of information due to integrating out one of the eigenvalues,
taking the unstable eigenvalue trajectory to be that
corresponding to the ZZ-boundary state \cite{ZZ} tensored with
Sen's boundary state \cite{sen-rolling}. This leads to the explicit
relation between the two different views in \cite{kms} and \cite{mgtv}
to compute decay amplitude
and reproduces
the expected time of decay
 $$
 \Delta t \sim \ln \hat \lambda\,.
 $$

Understanding the dynamics of the boundaries in the $2D$ string theory,
{\it i.e.}  the $D$-objects of the boundary Liouville
theory, from the RG analysis of general $(N \times N)$ matrices
on a circle is much more interesting.
In this case, by tuning the matrix
coupling constant and the mass parameter with $N \to \infty$ in
the double scaling limit, one tunes the bulk and boundary
cosmological constants respectively to arrive at a relation
between the two. This is given by the integration of the flow hitting the
$c=1$ fixed point. Such a relation presumably represent
Dirichlet or Neumann boundaries that are present in the Liouville
theory coupled to $c=1$ matter. We will show that for large circle
we arrive at such trajectories with correct scaling between the
two cosmological constants. However a detail analysis of such
trajectories for large and small circle will be extremely
interesting and will be studied in detail in a future publication.

The organization of the paper is as follows. In section 2, we
present the details of the RG calculation of matrix quantum
mechanics on a circle with a cubic potential and obtain the set of
beta function equations for the couplings. In section 3, we
analyze the flow, the fixed points and the critical exponents. As
understood previously, the usual KPZ-DDK scaling laws
\cite{KPZ,FD1,DK} for $c < 1$ are obeyed with a logarithmic
scaling violation for the $c=1$ model
\cite{GrossMiljkovic,BKZamo,GinsZ-J,Parisi,GK1}. With the
understanding of the RG flows, here we are able to obatin the
correct critical exponents of the $c=1$ strings, and to reproduce
the scaling behavior for the free energy with expected scaling
violation for the singlet sector. Also the RG trajectory near the
nontrivial fixed point shows the correct scaling behavior between
the bulk and boundary cosmological constants. From the running of
the prefactor of the partition function written in the
renormalized couplings, analogous to the running due to the wave
function renormalization, the free energy is observed to change
sign near $R=1$ for small value of the critical coupling. This is
reminiscent of the Kosterlitz-Thouless transition at self-dual
radius triggered by the liberation of the world-sheet vortices. In
section 4, we discuss that in what sense the RG analysis can
capture the T-duality respected by the singlet partition function.
In section 5, we discuss how the change of the world-sheet free
energy in integrating out one matrix eigenvalue can be seen in the
recent context to lead to the amplitude of the closed string
emission from the decaying brane. In section 6, we conclude with
some open questions.


\section{The Basic Set Up and the World-sheet RG Calculation}
\setcounter{equation}{0}

The existence of {\it double scaling limit} indicates that a
change in length scale induces flow in the coupling constants of
the theory in a way that one reaches the continuum limit with
desired critical exponents. In this continuum limit, as the matrix
coupling constant approaches a critical value $g_c$ the average
number of triangles $\langle n_G \rangle $ in triangulations at
any genus $G$ diverges with the exponent $-1$ while the length of the
triangles or the regularized spacing of the random lattice $a$ is
scaled to zero by taking $N \to \infty$ to keep the physical area
$a^2 \langle n_G \rangle$ or equivalently the string coupling
$g_s$ fixed,

\ba &&g \to g_c ~~\Rightarrow ~~\langle n_G \rangle \sim
(1-G)(\gamma_0-2)(1-g/g_c)^{-1} \to \infty \,,
\nonumber\\
&&N \to \infty ~~\Rightarrow~~a \sim N^{-\frac{1}{2-\gamma_0}} \to
0\,,
\nonumber\\
\mbox{with}~&&a^2 \langle n_G \rangle \sim
N^{-\frac{2}{2-\gamma_0}} (1-g/g_c) =
\mbox{const.}~~\mbox{or,}~~g_s^{-2} \equiv N^2
(g-g_c)^{2-\gamma_0}=\mbox{const}\,. \ea Thus one would naturally
try to understand how the two parameters of the theory, the size
of the matrices $N$ and the cosmological constant (mapped into the
matrix coupling $g$), evolve at the constant long distance physics
with the rescaling of the regularization length in the
triangulation of the world-sheet.  The flow equations will
automatically give rise to the correct scaling laws and the
critical exponents around the nontrivial $IR$ fixed points
governing the continuum physics. In the Wilsonian sense this can
be achieved by changing  $N\to N+\delta N$ by integrating out some
of the matrix elements, which is like integrating over the
momentum shell $\Lambda-d\Lambda < |p| < \Lambda$, and
compensating it by enlarging the space of the coupling constants
$g\to g+\delta g$. Here the space of coupling constants will
contain both the matrix coupling $g$ and the mass parameter $M^2$.

Following the RG scheme of Br\'ezin and Zinn-Justin we construct
the flow equations by integrating out a column and a row of an
$(N+1)\times (N+1)$ matrix, reducing it to an $N\times N$ matrix.
One expects the process to lead to the  following key relation
satisfied by the  matrix partition function

\be Z_{N+1}(g,M,R) = [\lambda(g,M,R)]^{N^2} Z_N(g',M',R')
\label{lambda-scaling} \ee with

\ba g' &=& g + \frac{1}{N}\beta(g,M,R)+O\Big(\frac{1}{N^2}\Big)\,,
\nonumber \\
M'^2 &=& M^2+\frac{1}{N}\beta(g,M,R)+O\Big(\frac{1}{N^2}\Big)\,,
 \label{beta-defn} \ea and

\be \lambda(g,M,R)=1+\frac{1}{N}r(g,M,R)+O\Big(\frac{1}{N^2}\Big)
\,. \label{r-defn} \ee Then the string partition function

\be \mathcal{F}(N,g,M,R)=\frac{1}{N^2}\ln Z_N(g,M,R)
\label{string-partition} \ee satisfies the Callan-Symanzik
equation

\be \Big[N\frac{\partial}{\partial
N}-\beta(g,M,R)\frac{\partial}{\partial
g}-\beta(g,M,R)\frac{\partial}{\partial M}+\gamma(g,M,R)\Big]
\mathcal{F}(N,g,M,R)=r(g,M,R) \,. \label{C-S} \ee


\subsection{Integrating out a row and a column}

Let us consider the $(N+1)\times (N+1)$ matrices $\phi_{N+1}(t)$
and decompose them into $N\times N$ matrices $\phi_N(t)$,
$N$-vectors $v_a(t)$ and $v_a^*(t)$ ($a=1,\ldots ,N$) and a scalar
$\alpha$. For the time being we can choose $\alpha=0$ as they are
of relative order $1/N$ and can be ignored in the double scaling
limit.

\be \phi_{N+1}(t) = \begin{pmatrix} \phi_N(t) &  v_a(t) \cr
v^*_a(t) & 0 \cr \end{pmatrix} \,. \label{matrixpara}\ee

For simplicity, we consider a cubic potential, which will serve as
a wall stabilizing the inverted oscillator potential. In terms of
$N+1$ dimensional matrix variable, the action reads

\be S_{N+1}[\phi_{N+1}(t),g,M,R] = (N+1)\int_0^{2\pi R} dt~\mbox
{Tr} \Big[\frac{1}{2}{\dot \phi}_{N+1}^2(t) +
\frac{1}{2}M^2\phi_{N+1}^2(t) -\frac{g}{3}\phi_{N+1}^3(t)\Big] \,.
\label{S1_{N+1}} \ee

The parametrization (\ref{matrixpara}) gives the simple relations,

\ba
&&\mbox{Tr}\phi_{N+1}^{2k} = \mbox{Tr}\phi_N^k+2k~v^*\phi_N^{2k-2} v+O(v^4)
\,,
\nonumber\\
 &&\mbox{Tr}\phi_{N+1}^{2k+1} = \mbox{Tr}\phi_N^k+(2k+1)~
 v^*\phi_N^{2k-1} v+O(v^4\phi^{2k-3}) \,.
\ea
The higher order terms in $v^*v$ can be neglected as they are supressed by
powers of
$O(1/N)$. The resulting partition function can be written as

\ba \mathcal{Z}_{N+1}[g,M,R] &=& \int_{\phi_N(2\pi R)=\phi_N(0)}
\mathcal{D}^{N^2}\phi_{N}(t) ~e^{-(N+1) \mbox{Tr} \int_0^{2\pi R}
dt~\{\frac{1}{2}{\dot \phi}_{N}^2(t) + \frac{1}{2}M^2\phi_{N}^2(t)
-\frac{g}{3}\phi_{N}^3(t)\}} \nonumber \\
& &\times \int_{v,v^*(2\pi
R)=v,v^*(0)}\mathcal{D}^Nv(t)\mathcal{D}^Nv^*(t)~e^
{-(N+1)\int_0^{2\pi R}dt~v^*(t)[-\partial_t^2+M^2 - g\phi_N
(t)]v(t)} \,.
\nonumber \\
& & \label{Z2_{N+1}} \ea

The above partition function is identical to the one considered in
\cite{yang-open,minahan-open}, where the $c=1$ matrix model
suitable for Veneziano type QCD has been considered to study open
strings. Both color $N$ and fermion (quark) flavor $N_f$ has been
taken to be large. Here the quarks are bosonic (the vectors). It
is precisely these fields in the fundamental representation of the
global $SU(N)$ group, which generate boundary terms in the Feynman
diagrams. As usual, we will use adiabatic treatment of first
integrating out the quark loops. Integrating over the quarks, we
get

\ba \mathcal{Z}_{N+1}[g,M,R] &=& \Big(\frac{\pi}{N+1}\Big)^N
\int_{\phi_N(2\pi R)=\phi_N(0)} \mathcal{D}^{N^2}\phi_{N}(t)
\nonumber \\ &&\times \exp\Big\{-\int_0^{2\pi R}
dt~\big[(N+1)\mbox{Tr}\Big\{\frac{1}{2}{\dot \phi}_{N}^2(t) +
\frac{1}{2}M^2\phi_{N}^2(t) -\frac{g}{3}\phi_{N}^3(t)\Big\}
\nonumber \\
&&+N~\mbox{Tr}\log\{-\partial_t^2+M^2 - g\phi_N (t)\}\big]\Big\}
\,.
\nonumber \\
& & \label{Z2_{N+1}-log} \ea

Logarithm with minus sign arise if the integration is performed on
$N$ flavors of fermions. Although such model has been considered
as matrix model for open strings, where the logarithmic term has
the effect of generating boundaries in the world-sheet, there are
some differences. Unlike the zero dimensional case
\cite{kaza-open,kostov-open,minahan-open-c=0}, this logarithmic
term does not arise by integrating out fields (quarks), which are
$1\times N$ matrices that couple to $\phi_N$. As a result, in
those open string models, the couplings in front of the logarithm
and in its argument are introduced by hand, rather than being
determined by the original closed string action. One tunes the
couplings to some appropriate values and the adiabatic treatment
of first integrating out the quark loops rise to tearing
phenomena. Interesting critical behavior of getting phases with
torn surface in the $c=0$ case \cite{kaza-open} also occur in the
one-dimensional case \cite{yang-open} if the dynamical loops are
generated by bosons without kinetic term. Absence of the kinetic
term makes fermion loops uncorrelated in time. One can ignore the
derivative term inside the logarithm if the mass and the couplings
are large enough. But there are interesting critical behavior when
the argument of the logarithm without the kinetic term approaches
zero \cite{boulatov-open,yang-open}. In \cite{minahan-open}, $c=1$
model with explicit expression for the fundamental fields has been
considered and is the one which is closest to our model obtained
after integrating out one row and one column of flavor degrees of
freedom. If one considers infinite line, only ground state is
relevant. For some choices of the coupling and the mass of the
particle moving around the boundary of the holes, it has been
possible to find the ground state \cite{minahan-open}, but the
exact spectrum is not known. We will return to this discussion in
section 5, when we will interpret the results from the RG analysis
in the recent context of time dependent tachyonic decay of
$D$-branes in two-dimensional string theory.

Now we return to evaluate the determinant by standard Feynman
expansion. In general it is a non-local object. For simplicity one
could consider the constant $\phi_N$ mode, which is equivalent to
studying the effective potential to determine the phase structure.
But here we will treat more general case of evaluating the
determinant with flavors coupling to general $\phi_N$ by
performing the calculation in Fourier transformed variables with
discrete momenta arising from the compact target space. The
complicated induced interactions arising from the logarithmic term
are ignored in small field approximation in order to get back the
beta functions that depend only on the original couplings.

Rescaling the vectors $v(t) \rightarrow \frac{v(t)}{\sqrt{2\pi
R(N+1)}}$, the $v$ dependent part of the partition function turns
out to be

\ba I[g,M,\phi_N,R]&=&\frac{1}{[\sqrt{2\pi
R(N+1)}]^{2N}}\int_{v,v^*(2\pi R)=v,v^*(0)}
\mathcal{D}^Nv(t)\mathcal{D}^Nv^*(t)
\nonumber \\
&&\exp\Big[-\int_0^{2\pi R}\frac{dt}{2\pi R}~\{\dot v^*\dot v
+v^*(M^2-g\phi_N)v \}\Big]\,, \ea
Before integrating out the vectors to get the determinant $\mbox{det}
||-\partial_t^2+M^2-g\phi_N||^{-1}$
and expanding it in Feynman diagrams, it is convenient
to Fourier transform all the fields as

\be
\mathcal{O}(t)=\sum_{m=-\infty}^{\infty}\mathcal{O}_m~e^{i\frac{m}{R}t}\,,~~
~
\mathcal{O}_m = \int_0^{2\pi R}\frac{dt}{2\pi
R}~e^{-i\frac{m}{R}t} \mathcal{O}(t)\,, \label{FourierTrfm} \ee
with
$$
\delta_{mn} = \int_0^{2\pi R}\frac{dt}{2\pi
R}~e^{i\frac{(n-m)}{R}t}\,,~~~ \delta(t-t') = \frac{1}{2\pi
R}\sum_{m=-\infty}^{\infty}e^{i\frac{m}{R}(t-t')} \,.
$$
In terms of the Fourier modes, the $v$-integration can be
expressed as

\ba I[g,M,\phi_N,R]&=&\frac{1}{[\sqrt{2\pi R(N+1)}]^{2N}}\int
\big(\prod_n dv_n^*dv_n^*\big) ~e^{-\sum_m
v_m^*\big(\frac{m^2}{R^2}+M^2\big)v_m
+g\sum_{m,l}v_m^*\phi_{m-l}v_l} \,,
\nonumber \\
\ea where we have neglected the $O(1/N)$ terms.

\subsection{One loop Feynman Diagrams}

\begin{figure}[htb]
\epsfysize=5cm \centerline{\epsffile{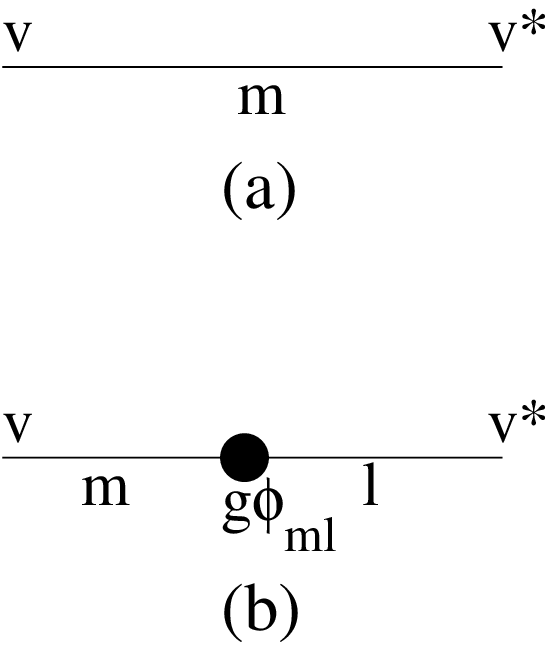}} \caption{The
propagators and vertices: (a) $[\mc{O}_{mn}^{v^*v}]^{-1}$, (b)
$[\mc{O}_{m-l}^{v^*v}(g,\phi)]^{-1}$.} \label{props}
\end{figure}

In order to carry out the $v$ integration diagrammatically, let us
now define the following operators

\ba \mc{O}_{mn}^{v^*v} &=&
\Big(\frac{mn}{R^2}+M^2\Big)\delta_{mn}\,,~~~
 \mc{O}_{m-l}^{v^*v}(g,\phi) = g\phi_{m-l} \,,
 \ea The inverse of these operators define various propagators and
vertices according to figure \ref{props}.

\begin{figure}[htb]
\epsfysize=3cm \centerline{\epsffile{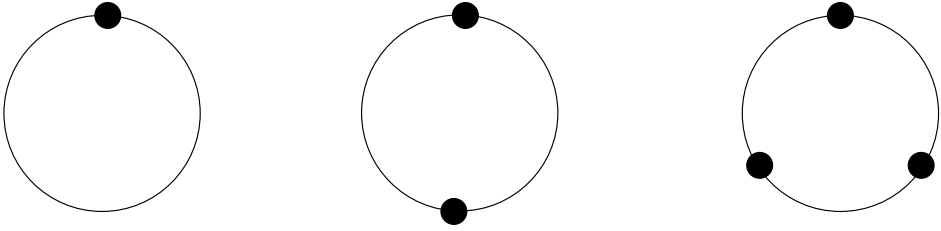}} \caption{The
diagrams contributing to $\Sigma$ at one-loop order.}
\label{diags}
\end{figure}

Hence the integral becomes

\be I[g,M,\phi_N,R,N] = \frac{1}{[2\pi
R(N+1)]^{N}}~\exp\Big[-\sum_{m_1,l_1}
\mc{O}_{m_1-l_1}^{v^*v}(g,\phi)\Big]~I_0(R,M,N)\,, \label{I1} \ee
where the gaussian part is as follows

\be I_0[R,M,N]=\int \Big(\prod_jdv_j^*dv_j\Big) ~\exp\big[-\sum_m
v_m^*\mc{O}_{mm}^{v^*v}v_m\big]\,. \ee In order to perform the
gaussian integration, we rescale $v_n$ as $v_n \to
v_n/(n^2/R^2+M^2)^{1/2}$ and use (\ref{FourierTrfm}). Performing
the gaussian integration, we get

\be I_0[R,M,N] = \Big(\frac{2\pi^4R^3}{\sinh^2\pi R M}\Big)^{N}
\prod_{n=1}^{\infty}\Big(\frac{R}{n}\Big)^{4N}\,, \ee where we
have used the standard relation

$$\prod_{n=1}^{\infty}\Big(1+\frac{x^2}{n^2}\Big)^{-1} = \frac{\pi
x}{\sinh\pi x}$$
Inserting this into (\ref{I1}), the $v$-integration becomes

\be I[g,M,\phi_N,R,N]=\mc{C}(R,M,N)~\Sigma[g,M,\phi_N,R,N]\,,
\label{I2} \ee where

\be \mc{C}(R,M,N) = \bigg[\frac{\pi^3R^2}{(N+1)\sinh^2\pi
RM}\bigg]^N \prod_{n=1}^{\infty}\Big(\frac{R}{n}\Big)^{4N}\,. \ee
In (\ref{I2}), $\Sigma[g,M,\phi_N,R,N]$ represents sum of one loop
Feynman diagrams as shown in figure \ref{diags}.

The sum $\Sigma [g,M,\phi_N,R,N]$ can be expressed as

\ba &&\Sigma[g,\phi_N,R,N] = 1 + g~\mbox{Tr}\bigg[\sum_n\frac{1}
{\frac{n^2}{R^2}+M^2}~\phi_0\bigg] 
\nonumber \\
&&+\frac{g^2}{2}~\mbox{Tr}\bigg[\sum_{m,n}\frac{1}{\big(\frac{m^2}{R^2}+M^2
\big)\big(\frac{n^2}{R^2}+M^2\big)}~\phi_{m-n}\phi_{n-m}\bigg]
\nonumber \\
&&+\frac{g^3}{3!}~\mbox{Tr}\Bigg[\sum_{m,n,l}\frac{1}
{\big(\frac{m^2}{R^2}+M^2\big)\big(\frac{n^2}{R^2}+M^2\big)
\big(\frac{l^2}{R^2}+M^2\big)}~\phi_{m-l}\phi_{l-n}\phi_{n-m}
\Bigg]+O(g^4) \label{SigmaFourier} \ea


\subsection{Evaluation of the diagrams}

In order to evaluate the one loop correction to the effective
action, we inverse transform the Fourier modes according to the
rule (\ref{FourierTrfm}) and sum-up the set of infinite series
using the formulae

\ba &&sum_{m=-\infty}^{\infty}
\frac{\exp[i(m/R)t]}{\frac{m^2}{R^2}+M^2} = \frac{\pi
R}{M}\frac{\cosh(\pi MR-Mt)}{\sinh\pi MR} \,,~~~0\le t\le 2\pi R \,,\ea
and,

\ba
\sum_{m=-\infty}^{\infty}\frac{m/R~\exp[i(m/R)t]}{\frac{m^2}{R^2}+a^2}
&=& \frac{i\pi R~\sinh(\pi a R - a t)}{\sinh \pi a R}\,,~~~ 0< t<
2\pi R \,,
\nonumber \\
&& i\pi R \,,~~~~~~~~~~~~~~~~~~~~~~~~~~t \to 0 \,,
\nonumber \\
&&-i\pi R \,,~~~~~~~~~~~~~~~~~~~~~~~t \to 2\pi R \,,
\nonumber \\
&& 0 \,,~~~~~~~~~~~~~~~~~~~~~~~~~~t=0=2\pi R\,.  \ea

These summations give the
corrections to the coefficients of the various terms in the
action, after $\Sigma[g,M,\phi_N,R,N]$ is exponentiated and
log-expanded using small field approximation.
Since after performing the inverse Fourier
transform, the various terms has nonlocal integrals over
several one dimensional dummy time variables, we breakup the
variables into center of mass and relative coordinates. Then we
expand the functions about the center of mass coordinates,
assuming the relative coordinates to be small enough, and consider
integration over the relative coordinates.

The details of the diagram evaluation are given in the appendix
taking into account all the above considerations and evaluating
the integrations over center of mass and relative time variables.
Collecting all the contributions up to $O(\phi\phi\phi)$ from the
appendix, the expression for $\Sigma[g,M,\phi_N,R,N]$ becomes

\ba \Sigma[g,M,\phi_N,R,N] &\simeq& 1+ F_{g1}(R,M)~g \int_0^{2\pi R}
dt~\mbox{Tr}\phi_N(t) + F_{g2}(R,M)~g^2 \int_0^{2\pi R}dt
~\mbox{Tr}\phi_N^2 (t)
\nonumber \\
&&+\hat{F}_{g2}(R,M)~g^2\int_0^{2\pi R}dt
~\frac{1}{2}\mbox{Tr}\dot\phi_N^2(t) +F_{g3}(R,M)~g^3\int_0^{2\pi
R}dt~\frac{1}{3}\mbox{Tr}\phi_N^3(t)\,.
\label{SigmaInvFourier}\nonumber
\ea
Note that, in the evaluation of $\Sigma$ we keep the contribution from the
nonlocal
terms of the action up to the kinetic term and the contribution from the
higher order terms
in the matrix field up to the cubic term. All other operators are redundant
for our purpose and
are negligible due to the small field approximation.

In the above expression, $F(R)$s are defined as follows

\ba F_{g1}(R,M)&&= \frac{1}{2M} \coth \pi MR\,,
\nonumber \\
F_{g2}(R,M)&&= \frac{1}
 {M^3 \sinh ^2 \pi MR }\Big( \frac{1}{2}\pi M R \cosh 2\pi MR+
 \frac{1}{8}\sinh 4 \pi
 MR \Big)\,,
 \nonumber\\
\hat F_{g2}(R,M) &&= \frac{1}{M^5 \sinh^2 \pi MR}\Big(\frac{\pi M
R}{16}\cosh 4\pi M R-\frac{\pi^3 M^3 R^3}{6} \cosh 2\pi MR
\nonumber\\
&&-\frac{1}{64}(1+8 \pi^2 M^2 R^2)\sinh 4\pi M R \Big) \,,
\nonumber\\
F_{g3}(R,M) &&=\frac{\pi M R}{64 M^5 \sinh^3\pi M R (\cosh 2 \pi M
R + \cosh 4 \pi M R ) }
\nonumber \\
&&[ 4 \pi M R \big(3 \cosh \pi M R + 2
\cosh3 \pi M R
+2 \cosh 5 \pi M R + \cosh 7 \pi M R \big)
\nonumber \\
&&+ \sinh \pi M R +
\sinh 3 \pi M R+ \sinh 5 \pi M R+ 2 \sinh 7 \pi M R
+ \sinh 9 \pi M R ]\,.
\nonumber \\
\label{defhyperbolic} \ea

 \subsection{Elimination of the tadpole term}

 The term proportional to $\int_0^{2\pi R}dt~\phi_N(t)$ is
 unwanted. As usual, in order to remove this term, we change
 the background $\phi_N(t) \to \phi_N(t)+f$, and set the
 net coefficient of the term linear in $\phi$ to zero.
 This fixes the value of $f$ as

\ba f &=&
\frac{1}{2}\Big(g+\frac{g}{N}+\frac{g^3F_{g3}}{N}\Big)^{-1}
\Big[\Big(M^2+\frac{M^2}{N} +\frac{g^2F_{g2}}{N}\Big) \pm
\Big\{\Big(M^2+\frac{M^2}{N}+\frac{g^2F_{g2}}{N}\Big)^2
\nonumber \\
&&+\frac{4gF_{g1}}{N}\Big(g+\frac{g}{N}+\frac{g^3F_{g3}}{N}\Big)\Big\}
^{\frac{1}{2}}\Big]~\simeq -g M^2 F_{g1}/N +
O\Big(\frac{1}{N^2}\Big)\,, \ea and accordingly modifies the
coefficients of all the terms in the action. After accommodating
all the changes,  the expression for $Z_{N+1}$ turns out to be

\ba Z_{N+1} &=& \mc{C}(R,M,N)~\exp[2\pi RN^2\mathscr{F}(g,M,R,N)]
\int\mc{D}^{N^2}\phi_N~\exp\Big[-N~\mbox{Tr}\int_0^{2\pi
R}dt~\Big\{\Big(1+\frac{1}{N}
\nonumber \\
&&+\frac{g^2\hat{F}_{g2}}{N}\Big)\frac{\dot\phi^2}{2}+\Big(1+\frac{M^2}{N}
+\frac{g^2F_{g2}}{N}-\frac{g^2M^2F_{g1}}{N}\Big)\frac{\phi^2(t)}{2}
\nonumber \\
&&-\Big(g+\frac{g}{N}+\frac{g^3F_{g3}}{N}\Big)\frac{\phi^3(t)}{3}\Big\}\Big]
\,,
\nonumber \\
\ea where the expression for $\mathscr{F}(g,M,R,N)$ is given by

\be \mathscr{F}(g,R,N) =
\frac{gF_{g1}}{N}f+\Big(M^2+\frac{M^2}{N}+\frac{g^2F_{g2}}{N}\Big)
\frac{f^2}{2}+\Big(g
+\frac{g}{N}+\frac{g^3F_{g3}}{N}{N}\Big)\frac{f^3}{3} \,, \ee
which is of $O\big(\frac{1}{N^2}\big)$.

\subsection{Rescaling of the fields and the variables}

In order to restore the original cut-off, we perform the following
rescalings so that the effective action is of the same form as the
bare one but with the renormalized strength of the couplings.

\be \phi_N(t) \to \rho\phi'_N(t')\,,~~~t' \to t(1-h~dl)\,,~~~R'
\to R(1-h~dl)\,,\label{rescaling}\ee where $dl=1/N$, and set the
overall coefficient of the kinetic term to zero. Here we can
assume that \be h=\sum_{i,j} c_{ij} g^i M^j h_{ij}(R)\,.\ee The
functional form of $h$ can be guessed from the contribution of the
Feynman diagrams in the behavior of the flow near the fixed
points. This sets the value of $\rho$ to \be \rho
=1+\frac{1}{2}(h-1+g^2 \hat F_{g2}) dl + O(dl^2),.\ee

\subsection{The beta function equations}

The effective action is of the same form as the bare one, but with
a renormalized strength of coupling. The resulting partition
function is given by

\ba Z_{N+1} &=& \lambda'~^{N^2}
\int\mc{D}^{N^2}\phi'_N(t')~\exp\Big[-N~\mbox{Tr}\int_0^{2\pi
R'}dt'\Big(\frac{\dot{\phi'_N}^2(t')}{2}+M'^2\frac{{\phi'_N}^2(t')}{2}
-g'\frac{{\phi'_N}^3(t')}{3}\Big)\Big] \,,\nonumber \\
\ea where,

\be \lambda'~^{N^2}= \mc{C}(R,M,N)~\exp[-2\pi
RN^2\mathscr{F}(g,M,N,R)]\rho^{N^2}\,. \ee Neglecting the
$O(dl^2)$ terms, the renormalized coupling and mass are expressed
in terms of the bare quantities as follows:

\ba g' &=& g+\Big(\frac{5}{2}h-\frac{1}{2}\Big)g~dl+\big[F_{g3}
(R,M)
-\frac{3}{2}\hat{F}_{g2}(R,M)\big]g^3~dl\,, \nonumber \\
M'^2 &=&
M^2+\big[2hM^2+g^2(1-M^2)F_{g2}(R,M)-g^2M^2\hat{F}_{g2}(R,M)\big]dl\,,
\nonumber\\
\lambda'&=& 1+\ln \Big[\frac{\pi^3 R^2}{\sinh \pi R M}\Big] dl
+\frac{1}{2}\big[(h-1)+g^2 \hat F_{g2}(R,M)\big]dl\,. \ea Here, in
simplifying the part $\mc{C}(R,M,N)^{\frac{1}{N}}$ in the
expression for $\lambda'$, we have assumed that for any value of
$R$, \ba \mc{C}(R,M,N)^{\frac{1}{N}}&&=\exp \Big[\frac{1}{N} \ln
\Big(\frac{\pi^3 R^2}{\sinh \pi R M}\Big) \Big]
\Big[\frac{\pi^3R^{4n}}{ (N+1)(n!)^4}\Big]^{\frac{1}{N}}_{n \to
\infty,~N \to \infty}
\nonumber\\
&& \simeq 1+\frac{1}{N} \ln \Big(\frac{\pi^3 R^2}{\sinh \pi R
M}\Big)+O(1/N^2)\,. \label{prefactor} \ea Also, the term
$\exp[-2\pi RN^2\mathscr{F}(g,M,N,R)]$ contributes only a factor
of $1$ as $\mathscr{F}(g,M,N,R)\sim O(dl^2)$. The resulting beta
function equations are given by

\ba \beta_g &=& \frac{d g}{dl} =
\Big(\frac{5}{2}h-\frac{1}{2}\Big)g + \big[F_{g3}(R,M)
-\frac{3}{2}\hat{F}_{g2}(R,M)\big]g^3 \,, \nonumber \\
\beta_{M^2} &=& \frac{dM^2}{dl} = 2hM^2 +g^2[(1-
M^2)F_{g2}(R,M)-M^2\hat{F}_{g2}(R,M)]\,, \nonumber\\
\beta_{\lambda} &=&\frac{d\lambda}{dl}=\ln \Big[\frac{\pi^3
R^2}{\sinh \pi R M}\Big] +\frac{1}{2}\big[(h-1)+g^2 \hat
F_{g2}(R,M)\big]\,.\label{bfe}\ea The relation (\ref{rescaling})
indicates in some sense a renormalization of the radius $R$ as \be
\beta_R = \frac{d R}{dl}= -h R\,.\label{betaR}\ee

Now before going to the detail analysis of the fixed points let us
try to understand few things about the structure of the beta
function equations. The much of the structure depends on
understanding the quantity $h$. One can clearly see for $g=0$ the
gaussian model is never expected to flow and hence for such a
fixed point $h=0$~({\it i.e.} corresponding to the trivial
rescaling $t'=t$ and $R'=R$). The situation is different for a
non-vanishing $h$. Demanding the mass parameter to be always at
some fixed value of $M$ (we will use $M^2=1$ for simplicity) and
$\beta_{M^2}=0$, one can easily determine some $h=h(R)$ for
nontrivial fixed points $g^* \ne 0$, as there are two independent
equations ($\beta_g=0, ~\beta_{M^2}=0$) and two unknowns ($h(R)$
and $g$). This is extremely interesting as it could indicate a
phase transition at certain radius due to turning on ({\it i.e.}
being relevant) the operator coupled to the cosmological constant.
Keeping $M^2=1$ for simplicity is consistent with the value of the
mass parameter ($M^2=\frac{1}{\alpha'}$) one originally works with
in the matrix partition function to visualize the matrix path
integral as the generator of the discretized version of the
Polyakov path integral of $2D$ bosonic string. In recent
identification of the matrix quantum mechanics with the quantum
mechanics of open string tachyon on unstable $D0$-branes the mass
parameter $M^2=\frac{1}{\alpha'}$ is identified with the open
string tachyon mass. However, we will use a framework of the flow
of a general $M^2$ ({\it i.e.} $h=0$~)to discuss the presence of
the boundaries.

Another point to be mentioned is that in this general analysis
though we will see the signature of the singlet sector to be more
apparent, the non-singlet sector does render indirect signature on
the flow. This is reflected in the $R$ dependence of the change of
the world sheet free energy with the scale, {\it i.e.} in the
$\beta_{\lambda}$. For very small  $g^*$ it exhibits a sign change
at certain radius  which is reminiscent of the tendency of the
non-singlet sector to be liberated at certain critical temperature
($R_c=1$). To capture this effect one needs to turn on the right
operator coupled to the fugacity of the vortices. In a separate
analysis we will work with a gauged matrix model with a gauge
breaking term that couples to the fugacity of vortices and drives
the system to the phase where the vortices are liberated, above a
critical temperature.

\section{The $c=1$ Fixed point}
\setcounter{equation}{0}

We will now analyze the fixed points of the flow equations given
by the simultaneous solutions of $\beta_g=\beta_{M^2}=0$. As we
discussed before, for the trivial rescaling of the coordinates and
the momenta, $t'=t$ and $1/R'=1/R$~, one can assume $h=0$. In this
case, the gaussian fixed point $\Lambda^{*}_{1}=\big(0,0\big)$
satisfies both the equations $\beta_g=0$ and $\beta_{M^2}=0$
trivially. The nontrivial fixed point $\Lambda^{*}_{2}$ is given
by, \be g^*= \pm \sqrt{ {\frac{1}{2F_{g3}(R,M^*)-3\hat
F_{g2}(R,M^*)}}}\label{gfp}\,,\ee where $M^*$ is determined by the
equation, \be g^{*2}\big((1-
M^{*2})F_{g2}(R,M^*)-M^{*2}\hat{F}_{g2}(R,M^*)\big)=0
\label{M-equation1}\,.\ee We will see in the next section that
this nontrivial value of $g^*$ will always characterize $c=1$
fixed point as it gives the critical exponent of $c=1$ for any $R$
for $h=0$. Using the value of $g^*$ at the nontrivial fixed point
(\ref{gfp}) in the equation (\ref{M-equation1}), one has \be
\frac{(1-
M^{*2})F_{g2}(R,M^*)-M^{*2}\hat{F}_{g2}(R,M^*)}{2F_{g3}(R,M^*)-3\hat
F_{g2}(R,M^*)}=0 \label{M-equation2}\ee For any value of $R$, this
equation trivially has solution for very small value of $M^*$.
However, for large $R$ there are also solutions for large $M^*$.
In the large $R$ limit the equation (\ref{M-equation2}) takes the
form \be M^{*2}\Big(\frac{8 \pi^2 M^{*2}R^2-8 M^{*2}-4 \pi M^{*}R+
9}{24 \pi^2 M^{*2}R^2-10 \pi M^{*}R+3}\Big)=0 \ee The solution for
large $M^{*}$ corresponds to the situation that the denominator is
much larger than the numerator (which solves with a precision of
$70 \%$ ). This gives \ba M^{*} \gg m_{+}\,,~~\mbox{or}~~M^{*} \ll
m_{-}\,,\nonumber\\\mbox{where}\,,~~ m_{\pm}=\frac{3 \pi R \pm
\sqrt{48+105 \pi^2 R^2}}{8(4 \pi^2 R^2+1)}\,.\ea As $R$ becomes
large, $m_{\pm}$ becomes smaller as $1/R$ and $M^{*}$ accesses all
possible values ranging from small to large. Thus the
nontrivial solutions of $\beta_g=0$ and $\beta_{M^2}=0$ gives dense
lines of fixed points occupying a region. Depending on all these
values of $M^*$, $g^* \ne 0$ will have different values from
(\ref{gfp}). Note that all these nontrivial fixed points
corresponds to $c=1$ as they have the string susceptibility
exponent of $c=1$~($\gamma_0=0$)~irrespective of the different
values of $g^*$ and $M^*$. It can be mentioned here that as $R \to
\infty$ the magnitude of $g^*$ becomes smaller and smaller and
eventually all the pair of fixed points for ~$\pm g^* \ne 0$~
coincides with the Gaussian fixed point.

Now let us look at the general shape of the trajectories flowing
to these $c=1$ nontrivial fixed points $g^* \ne 0$. The RG
equations $\beta_{g}=0$ and $\beta_{M^2}=0$ can be combined to
give \be \frac{\partial{M^2}}{\partial{g}}=\frac{2g
(F_{g2}-M^2(F_{g2}+\hat F_{g2}))}{-1+g^2(2 F_{g3}-3\hat
F_{g2})}\,,~~~~~(g \ne 0)\,. \ee  For large $R$ this equation
takes the following form where the variables are easily separable
\be \frac{\partial M}{\partial g} \sim \frac{
(F_{g2}-M^2(F_{g2}+\hat F_{g2}))}{g M(2 F_{g3}-3\hat F_{g2})}\ee
Using the explicit functional form at large $R$ this becomes \be
\frac{\partial M}{\partial g}\sim \frac{M(8 \pi^2 R^2 M^2-4 \pi R
M -8 M^2+ 9)}{g(24 \pi^2 R^2 M^2 -10 \pi R M +3)}\,.\ee The
leading behavior (considering 1/R as small parameter) indicates,
\be 3~\frac{dM}{M} \sim \frac{dg}{g}\,,~~~~~~\Rightarrow~~~~M^2
\sim g^{0.6}\label{Mscaling}\,. \ee In the last section we will
come back to this behavior in the context of Neumann and/or Dirichlet
boundaries of $2D$ string theory inserted by the integration of the
fundamental fields (quarks) $v^*$ and $v$. For Dirichlet boundaries
one expects large value of $M^*$. Then the boundary fields
become uncorrelated in
time. One can consider the scaling function of $M^2$ as the
boundary cosmological constant  $\mu_B$. Since $\mu_B$ is expected to
scale with the bulk cosmological constant $\mu$ as \be \mu_B \sim
\sqrt{\mu}\,, \ee and since the scaling function of $g$ gives the
renormalized bulk cosmological constant, the above scaling behavior in
(\ref{Mscaling}) is nothing but an indication of the presence of
various boundaries of the $2D$ string theory.

\subsection{The critical exponents }

Let us now go back to the matrix partition function we started
with. After completing the RG transformations, it obeys the
relation \be Z_{N+1}[g,M,R] \simeq
[\lambda(N,g,M,R)]^{N^2}~Z_N[g'=g+\delta g, M'=M+\delta
M,R'=R+\delta R] \,. \ee This leads to the Callan-Symanzik
equation \be \Big[N \frac{\partial}{\partial N}-\beta_g
\frac{\partial}{\partial g}-\beta_M \frac{\partial}{\partial
M}-\beta_R \frac{\partial}{\partial R}+\gamma \Big] \mc{F} [ g,M,R
] \approx  r [g, M, R] \,. \label{CS} \ee for the string partition
function (or the world-sheet free energy) \be \mc{F}[g,M,R] =
\frac{1}{N^2}\ln Z[g,M,R] \,, \ee with \be \gamma=2\,.\ee The
singular part of the world-sheet free energy $\mc{F}_s$ is given
by the solution of the homogeneous Callan-Symanzik equation. The
inhomogeneous part defined by the change in the prefactor
$\lambda$, contributes to subtleties in the free energy.

Let us  now discuss the critical exponents for the scaling
variables, the renormalized bulk cosmological constant
$\Delta=1-g/g^*$, the renormalized mass (or in the context of
boundaries, the renormalized boundary cosmological constant)
$\mc{M} = 1-M/M^*$. Introducing the matrix

\be \Omega_{k,l} = \frac {\partial \beta_{k}(\Lambda^*)} {\partial
\Lambda_{l} }\,, \ee the homogeneous part of the Callan-Symanzik
equation, satisfied by the singular part of the free energy, can
by rewritten as:

\be \Big[N \frac{\partial}{\partial N}-\Omega_{1} ~\Delta
\frac{\partial}{\partial \Delta} -\Omega_{2}~\mc{M}
\frac{\partial}{\partial~\mc{M}}+ h~ R \frac{\partial}{\partial R}
+2 \Big] \mc{F}_{s}~[~\Delta,\mc{M},R] = 0 \label{cs-scaling}\,,
\ee where, $\Omega_{i}$s are eigenvalues of the matrix
$\Omega_{k,l}$. They are nothing but the scaling dimensions of the
relevant operators. The general expressions of  $\Omega_{k,l}$s
and the eigenvalues $\Omega_1$, $\Omega_2 $, for different
fixed points (both Gaussian and the nontrivial one)  for a general
nonzero $h$ are evaluated in the Appendix B . For $h=0$, the
$\beta_R$ term drops out from the Callan Symanzik equation. The
singular behavior with respect to the renormalized cosmological
constant goes as, \be \mc{F} _{s} \sim \Delta ^{2/\Omega_{1}}
f_{1} [ N^{\Omega_{1}}~\Delta~]~f_{2} [ N^{\Omega_{2}}
~\mc{M}~]\,. \ee Comparing the above expression of $\mc{F}_{s}$
with the matrix model result ~$\mc{F}_{s} \sim \Delta
^{(2-\gamma_{0})} ~f[ N^{2/\gamma_{1}} \Delta]$, or using the
standard definition of the susceptibility ~$\Gamma ~\sim \frac
{\partial^2 \mc{F}_{s}} {\partial \Delta^2} \arrowvert_{~\mc{M}=0}
~\sim \Delta ^{-\gamma_{0}}~$, the string susceptibility exponent
$\gamma_{0}$ is given by

\be \gamma_{0} \sim (2-2/\Omega_{1})\,. \ee Note that in our
analysis $2/\gamma_1 \sim \Omega_{1}$, {\it i.e.} $\gamma_1 \sim
2/\Omega_{1}$ is consistent with the matrix model relation
$\gamma_0 + \gamma_1 = 2$. This relation is independent of the
explicit values of $\gamma_0$ and $\gamma_1$ and is easily
obtainable from the consideration of the torus. The string
susceptibility exponent at genus $G$ is defined by

\be \gamma_G = \gamma_0 + G~\gamma_1 \,.\ee  Referring to the
Appendix B, we observe that, for our nontrivial fixed point,
$\Omega_{1}=1-5h=1$ ($h=0$) , and hence $\gamma_{0}=0$. This shows
that our nontrivial fixed point is a $c=1$ fixed point at any $R$
for $h=0$. This $c=1$ nontrivial fixed points are repulsive with
respect to the flow of the parameter $g$ while the gaussian fixed
point is attractive. Similarly one can analyze the critical
exponents for the scaling function $\mathcal M$, the boundary
cosmological constant, which will characterize different boundary
conditions. Such an analysis will be useful to understand all
possible boundaries and to identify the Neumann and Dirichlet
case.

Note that in this RG, typically the nontrivial fixed point is
always situated close to the Gaussian fixed point. Hence,
$\Omega_{12},~ \Omega_{22}, ~\Omega_{21}~$ components of the
scaling dimension matrix are small. Thus $\Omega_{11} \sim
\Omega_1$ and $\Omega_2$ is also small.


\subsection{The logarithmic scaling violation of the singlet free energy}

In the previous section, the behavior $\mathcal {F}_s = \Delta^2$
of the singular part of the free energy near the $c=1$ nontrivial
fixed point ($\gamma_0=0$)  as a function of the renormalized
couplings or the scaling variables is consistent with the
continuum prediction, the KPZ-DDK scaling law~\cite{KPZ, FD1, DK}.
This power law dependence on $\Delta$ is present in all known $c <
1$ theories. However, for $c=1$, matrix model predicts a
logarithmic deviation to the usual power law
scaling~\cite{GrossMiljkovic,BKZamo,GinsZ-J,Parisi,GK1}. This is
known as the {\it logarithmic scaling violation} of the $c=1$
matrix model. In terms of the singlet free energy or the ground
state energy $E_0$, \ba &&- N^2 \mathcal{F}_s = E_0=-N^2 \pi^2
\frac{\Delta^2}{\ln \Delta}+ \ldots\,,
\nonumber\\
\mbox{or}\,,~~~ &&E_0=-\frac{N^2} {4 \pi} \mu^2 \ln \mu+
\ldots\,,~~~~~~\mbox{where}\,,~~~\Delta=-\frac{1}{2 \pi} \mu \ln
\mu\,. \ea In the inverse Laplace Transform with respect to the
area $A$, the sum over surfaces of fixed area $A$ behaves as \be
\mathcal{F} (A) \sim \frac{1}{A^3(\ln A)^2}\ee The continuum
methods seem to be insensitive to this peculiar dependence.

These scaling violations of the $c=1$ matrix model can be
naturally explained considering the lack of the translational
invariance in the Liouville dimension $\phi$ of the $2D$ string
theory, that is manifested by the $\phi$ dependence of the
background fields and is needed to maintain the conformal
invariance. Here, the Liouville dimension arises indirectly from
the semiclassical dynamics of the matrix eigenvalues while the
Euclidean time of the matrix quantum mechanics provides the other
dimension. The logarithmic factors in the expression of energy
actually reflects the logarithmic divergence of the Liouville
volume as the critical point is approached. The scaling violation
was argued to arise from the unusual dependence of the tachyon
potential on the Liouville field, $T(\phi)\sim \phi e^{\phi}$
~\cite{Polchinski-collective}. Later it was shown
that~\cite{kg9407014,kh9409064} considering interactions that
represent touching random surfaces give rise to new critical
behavior. This corresponds to the other branch where tachyon
potential has the ordinary Liouville form, $T(\phi)\sim e^{\phi}$,
with simpler scaling \be \mathcal{F}_s (\Delta) \sim \Delta^2 \ln
\Delta\,,~~~~\mathcal{F} (A) \sim \frac{1}{A^3}\,. \ee

We would now like to see how one can get this behavior peculiar to
$c=1$ from the Callan-Symanzik equation. The usual scaling
violation, $\mc{F}_s \sim \Delta^2/\ln \Delta$, directly follows
from the solution of Callan-Symanzik equation (\ref{CS}) 
near the $c=1$ nontrivial fixed point

$$
\bigg[N \frac{\partial}{\partial N}-\Delta
\frac{\partial}{\partial \Delta}\bigg] N^2 \mathcal{F}_s=0\,,
$$
up to a term that vanishes
in the scaling limit.

Also it is observed that the renormalization of $\lambda$ giving
rise to the right hand side of the inhomogeneous Callan-Symanzik
equation (\ref{CS}) can contribute logarithmic terms in the
behavior of the singlet free energy~\cite{fdavid}. The reason
right hand side is capable of contributing to the free energy
(universal physics) is that it actually measures the change of the
world sheet free energy with the change of scale and can be
expressed in the same footing as the terms in the left hand side
of the Callan-Symanzik equation \be \beta_{\lambda}\frac{\partial
\mathcal{F}}{\partial \lambda}=\frac{\partial\mathcal{F}}{\partial
(\frac{1}{N})}\,. \ee Thus, using our Callan-Symanzik equation
~(\ref{CS})~ and the linearized beta function equations
~(\ref{bfe})~around the nontrivial fixed point~(\ref{gfp})~ we
have~, \be \bigg[N \frac{\partial}{\partial N}-\Delta
\frac{\partial}{\partial \Delta}\bigg] N^2
\mathcal{F}=\beta_{\lambda}\frac{\partial (N^2
\mathcal{F})}{\partial \lambda}=-\Delta \hat F_{g3} (R) g^*+
\Delta^2 \frac{\hat F_{g3}}{2} \ee This gives \ba 2
\mathcal{F}-\Delta \frac{\partial \mathcal{F}}{\partial \Delta}
=N^{-2}\Big(-\Delta \hat F_{g3} (R) g^*+ \Delta^2 \frac{\hat
F_{g3}}{2}\Big)\,,
\nonumber\\
\Rightarrow~~~~ \partial(\mathcal{F}/\Delta^2)
=N^{-2}\Big(\frac{\hat F_{g3} (R)}{2 \Delta}- \frac{\hat F_{g3}
(R) g^*}{\Delta^2}\Big) \partial \Delta\,.
\nonumber\\
\ea Integrating both sides gives rise to the scaling \be E_0=-N^2
\mathcal{F} \sim -\frac{1}{2} \hat F_{g3} (R)~ \Delta^2 \ln \Delta
\,. \ee However, this dependence does not give rise to the
logarithm in the inverse Laplace transform with respect to the
area. This is rather the behavior corresponding to the other
branch of the tachyon of the Liouville theory. Perhaps this
reflects the fact that even though we are not explicitly
considering the branching interactions representing the touching
random surfaces generated by the redundant higher order terms,
they have indirect effect in the change of the world sheet free
energy.

Note that in this analysis we are not looking at the $\mathcal M$
dependence of the free energy since typically $\Omega_2$ is very
small and hence can be ignored in ~(\ref{CS})~or
in~(\ref{cs-scaling})~. Alternatively, one can also consider a
fixed mass parameter $M$ in doing the analysis.

\subsection{The Radius Dependence of the free energy}

As we have discussed before in context of the analysis of the
structure of the beta function equations, one can observe an
indirect impact of the world sheet vortices on the free energy in
the sense that it flips its sign at the self-dual radius $R=1$.
This radius is known to be the inverse temperature for the
Kosterlitz-Thouless phase transition due to the liberation of the
world-sheet vortices~\cite{GK2}. To observe this we would demand
that the mass parameter $M$ is fixed to a certain value (for
simplicity we choose $M=1$) while $\beta_{M^2}$ is set to zero.
Then for nontrivial fixed points of $\beta_{g}=0$, the parameter
$h$ can be determined as a function of $R$ by solving the two
simultaneous equations $\beta_g=0$ and $\beta_{M^2}|_{M=1}=0$ for
the two unknowns ~$h(R)$~ and ~$g^*$~. This gives, \be
h(R)=\frac{1}{\Big(4 ~F_{g3}(R)/\hat
F_{g2}(R)-1\Big)}\,.\label{KTSDR} \ee In low $R$ approximation
$h(R) \sim 1/R^2$ and thus the change of the world-sheet free
energy around the nontrivial fixed point becomes (considering
$g^*$ to be very small) \be \beta_{\lambda} \approx
\frac{1}{2}(1/R^2-1) \,. \ee This indicates a phase transition at
$R=1$.

\begin{figure}[htb]
\epsfysize=6cm \centerline{\epsffile{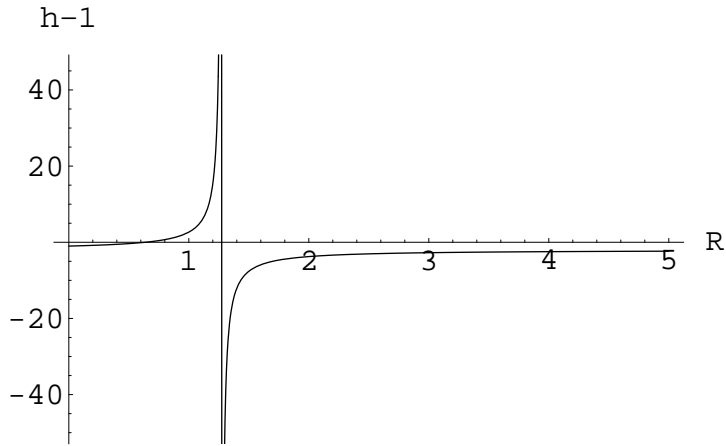}} \caption{Behavior
of $h(R)-1$ at relatively large $R$.} \label{h(R)}
\end{figure}
However, a plot of ~$\beta_{\lambda}\approx \frac{1}{2}(h(R)-1)$~
using a simplified $h(R)$ (~$h(R) \sim \frac{1}{-1+1.27/R}$~) in
relatively large $R$ expansion shows the discontinuity at
~$R=1.2$~(Figure \ref{h(R)}). In this discussion the logarithmic
constant in the equation for $\beta_{\lambda}$~(\ref{bfe})~is
ignored as it merely indicates some vacuum normalization and can
be conveniently absorbed. Another important point to be mentioned
is that the high~ $R$~ approximation for $h(R)$ is not quite
efficient to compute the correct exponent of the $c=1$ fixed point
(in the sense that $h(R)$ does not go to zero for large $R$
instead approaches $-1$ in that limit). Since this is purely
dependent on the contribution of the single line Feynman diagrams
in integrating out the quark loops, we need further improvement in
the computation of the diagrams.

\section{The T-duality of the singlet partition function}
\setcounter{equation}{0}

In this section we would analyze that in what sense our RG can
capture the T-duality respected by the singlet sector of the MQM.
Let us consider the matrix partition function in the eigenvalue
representation by diagonalizing the matrices by $SU(N)$
transformation and suitably integrating out the angular degrees of
freedom. The diagonalization of $\Phi(t)$ gives,

\be \Phi_{i j}(t)=\sum_{k=1}^{N}  \Omega_{i k}^{\dagger}(t)
\lambda_{k} (t) \Omega_{k j} (t)\,. \ee Hence,

\be \mbox{Tr} ~\dot {\Phi}^2=\sum_{k=1}^{N} \dot {\lambda}_{k}^2 +
\sum_{k \neq j} (\lambda_{k}-\lambda_{j})^2  |A_{k j}|^2\,,~~A_{k
j}(t)= ( \Omega^{\dagger}(t) \Omega(t) )_{k j}\,. \ee Here the
gauge field $A$ acts as a lagrange multiplier that projects onto
$SU(N)$ singlet wave functions which depend on the $N$ matrix
eigenvalues only.  In the recent works on $N \times N$ matrix
quantum mechanics as the quantum mechanics of string tachyon field
on the $N$ unstable  $D0$-branes, this gauge field is identified
as the non-dynamical gauge field in the open string spectrum
corresponding to the vertex operator $\dot{t}$. Due to the
periodic boundary condition on the $\Phi(t)$ and on the
$\Omega(t)$-s the gauge field $A_{ij}(t)$-s are constrained:

\be \hat{T}\exp~i\int_0^{2\pi R} A(t)~dt = I\,. \ee Taking into
account of the above constraint (which actually projects to the
singlet sector) in the measure of the path integral, the (singlet)
partition function can be formally written as (see the
review~\cite{kazakov}):

\be Z_{N}(g,R)=\sum_{r} d_{r}~d_{r}^{(0)} \int \prod_{k} \mc{D}
\lambda_{k} (t) e^{-\int _{0}^{2 \pi R} dt [N \sum_{i}^{N}
(\frac{1}{2} \dot \lambda_{i}^2+V(\lambda_{i}))+\frac
{C_{r}^{(2)}}{2 N^3} \sum_{i \neq
j}\frac{1}{(\lambda_{i}-\lambda_{j})^2}]}\,, \ee where, the
$d_{R}$ and $d_{R}^{(0)}$ are the dimensions of the $R$-th
representation of $U(N)$ and that of the subspace of vectors with
the zero weights in $R$ and $C_{r}^{(2)}$ is the value of the
quadratic Casimir operator in the $r$-th representation.

\be C_{r}^{(2)} \simeq  N\,. \ee

Note that, in this process, all the Vandermonde determinants are
gone but the fermionic statistics is still maintained for the
eigenvalues are now periodic:

\be \lambda_{i}(0)=\lambda_{i}(2 \pi R)\,,~~~i=1,...,N \,. \ee In
a similar sense to our scheme, let us now begin with the partition
function of $(N+1)$ eigenvalues and observe the effect of
integrating out one of the eigenvalues in presence of the rest of
the $N$ eigenvalues. Physically this would mean integrating out
one of the fermions in the presence of the interaction due to the
rest of the $N$ fermions. Then the Fermi sea readjusts it's height
and this renormalizes the parameters of the theory like the
cosmological constant. In recent works a similar picture has been
discussed in context of closed string radiation due to the rolling
open string tachyon on unstable $D0$ branes represented by the
unstable matrix eigenvalues. We will come to that in the next
chapter.

Now the partition function of $(N+1)$ eigenvalues can be split as,
\ba Z_{N+1}(g,R)=\sum_{r} d_{r}~d_{r}^{(0)} \int \prod_{k=1}^{N}
\mc{D} \lambda_{k} (t) e^{-\int _{0}^{2 \pi R} dt \big [ (N+1)
\sum_{i}^{N} (\frac{1}{2} \dot \lambda_{i}^2+
\frac{1}{2}\lambda_{i}^2-\frac{g}{3} \lambda_{i}^3) +\frac {1}{2
(N+1)^2} \sum_{i \neq j}\frac{1}{(\lambda_{i}-\lambda_{j})^2}
\big]}
\nonumber\\
\int \mc{D} \lambda_{N+1} (t) e^{-\int _{0}^{2 \pi R} dt \big[
(N+1) (\frac{1}{2} \dot \lambda_{N+1}^2+\frac{1}{2}\lambda_{N+1}^2
-\frac{g}{3} \lambda_{N+1}^3)+\frac {1}{2 (N+1)^2} \sum_{j=1}^{N}
\frac{1}{(\lambda_{N+1}-\lambda_{j})^2}\big]}\,. \label{ffrg1} \ea
Let us consider the simplest form of real (Hermitian Matrix model)
and periodic eigenvalue, \be \lambda_{N+1}(t)=\lambda_{N+1}
\cos{t/R}\,. \ee Then the measure over the $(N+1)$-th eigenvalue
trajectory (to be integrated out) becomes, \be \mc{D}
\lambda_{N+1} (t) =\prod_{k=1}^{N} d\lambda_{N+1} \cos(2 \pi k/N)=
\prod_{k=1}^{N} d\lambda_{N+1}(1-O(1/N^2)) \simeq \prod_{k=1}^{N}
d\lambda_{N+1}\,. \ee Hence the integration over $\lambda_{N+1}$
gives, \ba I_{N+1}&&=\prod_{k=1}^{N} \int
d\lambda_{N+1}~e^{-(N+1)\Big[ \frac{1}{2}
\lambda_{N+1}^2\int_{0}^{2 \pi R} dt \big(\frac{1}{R^2}
\sin^2{t/R}+\cos^2{t/R}\big)-\frac{g}{3}
\lambda_{N+1}^3\int_{0}^{2 \pi R} dt \cos^3{t/R}\Big] }
\nonumber\\
&&~~~~~~~~~~~e^{-\frac {1}{2 (N+1)^2} \big[ \sum_{j=1}^{N}
\frac{1}{(\lambda_{N+1}-\lambda_{j})^2}\int_{0}^{2 \pi R}\frac{dt
}{\cos^2{t/R}}\big]}
\nonumber\\
&&= \prod_{k=1}^{N} \int d\lambda_{N+1}~e^{-\frac{\pi}{2}
(N+1)(R+1/R) \lambda_{N+1}^2} =
\big(2/(N+1)(R+1/R)\big)^{\frac{N}{2}}\,. \ea As before,
considering the rescaling of the fields \be \lambda_{i}(t) \to
\rho \lambda_{i}'(t)\,,~~~i=1,..,N\,, \ee and setting the
coefficient of $\frac{1}{2}\dot \lambda_{i}^{' 2}(t)$ to one
(which in this simplest case simultaneously sets the coefficient
of $\frac{1}{2}\lambda_{i}^{' 2}(t)$ to one too, and the mass term
does not run with the scale change) we have,

\ba \rho=\big(1+\frac{1}{N}\big)^{-\frac{1}{2}} \simeq \big(
1-\frac{1}{2 N} + O(1/N^2) \big) \,, \ea Hence, the partition
function can be rewritten as, \ba
Z_{N+1}(g,R)&&=\Big(\frac{2}{(N+1)(R+1/R)}\Big)^{\frac{N}{2}}
\big( 1-\frac{1}{2 N} + O(1/N^2) \big)^{N}
\nonumber\\
&& \sum_{r} d_{r}~d_{r}^{(0)} \int \prod_{k=1}^{N} \mc{D}
\lambda_{k}' (t) e^{-\int _{0}^{2 \pi R} dt \big [N \sum_{i}^{N}
(\frac{1}{2} \dot \lambda_{i}^{' 2}+ \frac{1}{2}\lambda_{i}^{'
2}-\frac{g'}{3} \lambda_{i}^{' 3})+\frac {C' }{2 N^2} \sum_{i \neq
j}\frac{1}{(\lambda_{i}'-\lambda_{j}')^2} \big]}\,,
\nonumber\\
&&g'=g-\frac{g}{2 N}+O(\frac{1}{N^2})\,,~~~~C'=
1-\frac{1}{N}+O(\frac{1}{N^2})\,. \label{ffrg2} \ea This leads to,
\ba \frac{Z_{N+1}(g,R)}{Z_{N}(g',R)}&&=\Big(\frac{2}
{(N+1)(R+1/R)}\Big)^{\frac{N}{2}}\big(1-\frac{1}{2 N} + O(1/N^2)
\big)^{N}\,,
\nonumber\\
&&=\Big[\frac{1}{N^{\frac{1}{2N}}}
\Big(\frac{2}{R+1/R}\Big)^{\frac{1}{2 N}} (1-1/N)^{\frac{1}{N}}
\Big]^{N^2}\,. \label{ffcs} \ea Note that, due to the cosine,
which is the simplest choice for the eigenvalue $\lambda_{N+1}$,
both the interaction terms corresponding to the cubic
self-interaction of $\lambda_{N+1}$ and the mutual (repulsive
coulomb) interaction with the rest of the $N$ eigenvalues drop out
in the integration $I_{N+1}$. As a result in the equation
(\ref{ffrg1}), the integration over the $N+1$-th eigenvalue
becomes just a overall prefactor (contributing to the change of
the world-sheet free energy associated with the readjustment of
the Fermi level  in response to loosing one of the fermions) as in
(\ref{ffrg2}), unlike the more general situations studied in the
previous chapters where the integration over a part of the matrix
degrees of freedom does produce nontrivial one loop correction
terms that adds up to the left over $N$-fermion partition
$Z_{N}(g',R)$ besides producing the overall prefactor. Even in
simple eigenvalue representation, a more general choice of the
functional form of $\lambda_{N+1}$ would produce such correction
terms. In those cases the beta function for $g$, $\beta_g =
\frac{\delta g}{\delta (\frac{1}{N})}= -\frac{g}{2}+...$, will
have higher order terms in $g$ (which can give nontrivial fixed
points). Note that the leading term in the beta function,
$-\frac{g}{2}$, is the same in any case.

 Now the equation (\ref{ffcs}) implies the Callan-Symanzik
 equation for the world sheet free energy
 $\mc{F} [ g,R]= \frac{1}{N^2}\ln Z[g,R] $:
 \be
\Big[N \frac{\partial}{\partial N}-\beta_g
\frac{\partial}{\partial g}+\gamma \Big] \mc{F} [ g,R] \simeq
\frac{1}{2}\ln \big(\frac{2} {R+1/R}\big)+O(1/N)\,,~~\beta_g
\simeq -\frac{g}{2}\,,~~~\gamma=2\,.
 \ee
 In terms of the scaling variable $\Delta=(1-g/g*)$, the Callan-Symanzik
 equation in the Double scaling limit is given by,
 \be
\Big[N \frac{\partial}{\partial N}-\frac{1}{2}\Delta
\frac{\partial}{\partial \Delta}+\gamma \Big] ~\mc{F} [ \Delta,R]
~\simeq ~\frac{1}{2} \ln \big(\frac{2} {R+1/R}\big)\,.
 \ee
Solution of the homogeneous equation is same as before.  Using the
same ansatz as before, $\mc{F}[\Delta,
R]=N^{-2}f_{1}(\Delta)f_{2}(R)$, and solving the inhomogeneous
part we observe that the world sheet free energy has a functional
form of $f(R+1/R)$ with respect to $R$: \be \mc{F}[\Delta, R] =
-\ln \Delta~\ln \big(\frac{2}{R+1/R}\big)\,. \ee This shows that
the large $N$ RG is capable of  capturing the {\it T-duality}
property respected by the singlet partition function. However, two
comments are note worthy here. First of all, in this simplest
case, the beta function does not receive higher order corrections
in the coupling constant. Hence the fixed point is a trivial
Gaussian fixed point with inadequate critical exponent
($\Omega_{1}= -\frac{1}{2}, \gamma_{0}=2-2/\Omega_{1}=6$) to
describe the $c=1$ fixed point ($\gamma_{0}=0$). Actually, from
the results of the more general set up in the previous chapters,
this Gaussian fixed point could be thought of to be overlapping
with a pair of double zero of the beta function at infinitesimal
distance from it for large $R$ (where the singlet free fermion
picture is meaningful). With a more general choice of the
functional form of the eigenvalue, the gaussian fixed point will
be resolved into a nontrivial double zero of the beta function
with appropriate critical index to describe $c=1$, as obtained in
the previous chapters. This $c=1$ fixed point would explicitly
show {\it T-duality} in exactly the same manner from the gaussian
part of the integration over the $N+1$-th eigenvalue in presence
of the rest of the $N$ eigenvalues.  Secondly, we note that the 
T-duality arises in the usual sense from the kinetic and 
the quadratic term of the action in the wave function of the 
$N+1$-th fermion. This wave function can be compared with the 
free fermion wave functions for the singlet sector of $c=1$
on large circle. A small observation is that in  the more general 
set up (with $M=1$ as in here), the
gaussian part of the integration over a part of the matrix degrees
of freedom produces the functional form $f(\sinh \pi R /\pi^2 R^2)
\simeq f(\frac{ 1}{\pi R}+\frac{ \pi R}{3}+O(R^2))$  (to compare
see equations  (\ref{prefactor})),
instead of producing $f(R+1/R)$. This shows how the T-duality is
broken at this level.

\section{Comments on D-brane Decay and Rolling Tachyon}
\setcounter{equation}{0}

The recent observations \cite{mgv,kms,mgtv,martinec} show that the
matrix model can be realized as the effective dynamics of
$D$-branes in $c=1$ non-critical string theory. We briefly review
the basic picture before going into the matrix model RG
interpretation of the open string rolling tachyon in string
theory.

\subsection{Review of D-brane Decay in 2D String Theory}

According to this conjecture, the matrix $\Phi_N$ itself can be
seen as an open string tachyon field on $D0$-branes and the matrix
potential as the tachyon potential. The $SU(N)$ symmetric matrix
quantum mechanics in an upside down inverted harmonic oscillator
potential, {\it i.e.} in the double scaling limit, exactly
describes open strings on $N$ $D0$-branes and is dual to Liouville
theory coupled to $c=1$ matter, that describes the two dimensional
closed string theory together with its $D0$-branes. The
$D0$-branes, on which the tachyon field resides, are described by
the localized boundary states for the Liouville coordinate $\phi$
introduced by A.B. and Al.B. Zamolodchikov \cite{ZZ} tensored with
Sen's rolling tachyon boundary states with Neumann boundary
condition for the free time direction $t$ \cite{sen-rolling}. They
are localized in the strong coupling region $\phi\to \infty$ far
from the bulk region of large and negative $\phi$ (signalled by
the absence of "bulk poles" in the one-point function of the
closed string vertex operator on the disc). These branes are
parametrized by the pair of integers $(m,n)$. Among them only the
set $(1,n)$ has smooth behavior with no singularity in the
classical limit $b \to 0$, where $b$ measures the "rigidity" of
the $2D$ surface to quantum fluctuations of the metric. The one
labelled as $(1,1)$ contains operator that matches with the
massless tachyon in the open string spectrum and is the only one
which has been found to be consistent with the standard loop
perturbation theory and consequently is identified as the boundary
state at one-dimensional infinity or the "absolute" of Euclidean
$AdS_2$ ({\it i.e.} classical Lobachevskiy plane) \cite{ZZ}. It is
still not clear what the other boundary states with $m\ne 1$,
$n\ne 1$ correspond to.

There is also another class of Liouville boundary
states~\cite{FZZ}, extended in the bulk weak coupling region of
$\phi \to -\infty$ (exhibited by the poles in the bulk one point
function). They correspond to $D$-strings when the time direction
is taken to be Neumann. The continuous families of such boundary
states are parametrized by the uniformization parameter $s$, given
by

\be \mu_B = \sqrt{\mu}\cosh \pi s \,. \label{uniformization}\ee
Here $\mu_B$ is the renormalized boundary cosmological constant
and $\mu$ is the renormalized bulk cosmological constant in the
Liouville theory:

\ba \mu_B &\equiv&  (\pi \mu_{B0} \gamma (b^2))\,,
\nonumber\\
\mu &\equiv& (\pi \mu_0 \gamma(b^2))\,,
\nonumber\\
\gamma (b^2) &=& \frac{\Gamma(b^2)}{\Gamma(1-b^2)}\,, \ea in the
sense that, in order to get finite amplitude, they are kept finite
in the limit $b \to 1$, {\it i.e.} $c_L \to 25$. All such
$D$-branes in this family have a continuous spectrum of open
string states, namely, the massless open string tachyons. For the
choice of $s=\frac{i}{2}(2n+1)$, $m \in \mathbb{Z}$, one can set
$\mu_B=0$. The work of \cite{mgv} uses the minimal values $s=\pm
\frac{i}{2}$. Increasing $m$ gives rise to increasingly tachyonic
open string modes.

As the tachyon is rolling down the potential, one can identify the
boundary cosmological constant $\mu_B$ of the Liouville theory
with the time dependent tachyon potential $\lambda \cosh t$ as
follows:

 \be
 \mu_B(t) = \sqrt{\mu}\cosh s(t)=\lambda \cosh t\,.
 \ee
As $t \to -\infty$, $\mu_B \to 0$, which corresponds to unstable
$D$-string picture. As $t \to \infty$, $\mu_B$ and $s$ grow to
large values. In this process, the unstable $D$-string one starts
with, decays into closed string vacuum of Liouville theory with
its localized $D0$-branes and all the open string excitations on
the $D$-string are pushed to $\phi \to -\infty$. In the matrix
model picture, an unstable brane corresponds to a free fermion on
the top of the inverted harmonic oscillator potential rolling down
as an unstable eigenvalue $z(t)=\lambda e^t$ from the top to the
Fermi level $\mu_F$.

The amplitude for such $D$-brane tachyonic decay with momentum $P$
and energy $\omega=|P|$ is given by the expectation value of a
normalizable bulk operator $v_{\omega ,P} = \exp
[(2+iP)\phi+i|P|t]$ inside the disk. As was shown in \cite{kms}
this disk one-point function agrees with the outgoing radiation
amplitude derived from $c=1$ matrix model (taking into account the
leg-pole factors). The disk one-point function in the two
dimensional string theory is given as follows:

\be A (\omega , P) = A_t(\omega)A_L(P) = \langle
v_\omega|B_{\hat\lambda}\rangle\langle v_P|B_s\rangle\,,\ee where
the two factors denote the time part (obtained in
\cite{LiuLambMald}) and the Liouville part (obtained in \cite{FZZ}
and \cite{ZZ} for the two kinds of branes) respectively. Although
the end result of tachyonic decay for both the branes are the same
and the time part of the amplitude is identical, the Liouville
part or the bulk one point functions for the two cases differs
with the absence of pole for the localized brane.

The time part of the amplitude $A_t(\omega) = \langle
v_\omega|B_{\hat\lambda}\rangle$ depends on the choice of contour
\cite{LiuLambMald} and is described by the boundary
interaction\footnote{Here $\hat\lambda = \sin \pi\tilde\lambda$ in
order to match with the trajectory considered in
\cite{sen-rolling}.} $\hat\lambda \cos t/\alpha'$. Note that, in
our convention, $t$ denotes the Euclidean time. In the Lorentzian
coordinate $X^0=it$, the boundary interaction is given by
$\hat\lambda \cosh X^0/\alpha'$. The choice of Hartle-Hawking
contour gives rise to the following result \cite{LiuLambMald}:

\be A_t(\omega) = \langle \exp[i\omega t] \rangle =
\frac{\pi\exp[-i\sqrt{\alpha'}\omega\log\hat\lambda]}{\sinh
\pi\sqrt{\alpha'}\omega}\,,\ee having pole at $\omega = 0$,
signaling the presence of continuous spectrum in the on-shell open
string channel \cite{PonsotSchomerusTeschner,mgtv}.

The Liouville part of the amplitude for $D1$-brane decay (for the
choice of $s=\pm i/2$) can be evaluated from the on-shell CFT
calculation of the boundary state constructed by Fateev,
Zamolodchikov and Zamolodchikov \cite{FZZ} resulting the total
amplitude to be

\be \mathcal{A}_{D 1}(\omega, P)= 2\pi^2~
e^{-i\delta(P)}~e^{-i\sqrt{\alpha'}|P|\log\hat\lambda}
~\frac{\cos\pi\sqrt{\alpha'}s
P}{\sinh\pi\sqrt{\alpha'}\omega~\sinh\pi\sqrt{\alpha'}P}  \,. \ee
The corresponding phase factor is given by

\be e^{-i\delta(P)} =
i\mu^{-i\sqrt{\alpha'}P/2}~\frac{\Gamma(i\sqrt{\alpha'}P)}
{\Gamma(-i\sqrt{\alpha'}P)}\,. \ee

The corresponding disk one-point function  for the $D0$-brane has
no pole \cite{kms}. The Liouville part is described by the
Zamolodchikov boundary state \cite{ZZ} and the total amplitude is
given by

\be A_{D0}(\omega, P) =
-\mbox{sgn}(P)\pi~e^{-i\delta(P)}~e^{-i\sqrt{\alpha'}|P|\log\hat\lambda}
\,.\ee

Let us now try to understand this from matrix quantum mechanics
picture. The physics of the matrix model in eigenvalue
representation is that of $N$ non-interacting fermions, moving in
the potential $V(\lambda)$. In the double scaling limit, the
potential is that of an inverted harmonic oscillator. Each fermion
occupies a volume of $2\pi/\beta =2\pi g^2/N$ in phase space,
where $1/\beta$ plays the role of $\hbar$. All levels with
single-particle energy $\frac{1}{2}v^2+V < \mu_F$ are filled. As
$g$ increases or $\beta$ decreases, the phase space volume
occupied by the fermionic states grow. At the critical value, $g
\to g_c$, the fermi level reaches the top of the potential, $\mu_F
\to \mu_c$, where the fermions start to spill over the barrier.
The free energy is singular at this point. In the double scaling
limit, $N\to\infty$ and $g\to g_c$, holding the string coupling
$g_s \equiv \bar\mu^{-1} = [\beta(\mu_c-\mu_F)]^{-1}$
fixed\footnote{In terms of $\Delta = g-g_c \simeq \mu\ln\mu$, we
have kept $g_s = [N\Delta^{(2-\gamma_0)/2}]^{-1}
=(N\Delta^{1/\Omega_{11}})^{-1}$ fixed in our RG analysis in
previous sections.}. In other words, the ratio of the gap between
fermionic levels (which is of order $\hbar \sim 1/\beta$) to $\mu
= \mu_c - \mu_F$ remains constant in the double scaling limit. In
this limit the physics is that of the Fermi sea in inverted
harmonic oscillator with the hamiltonian $H(v,\lambda) =
(v^2-\lambda^2)/2+\bar\mu$, where $\lambda$, $v$ are canonical
conjugate variables. The collective motions of the fermions are
classically described in terms of a time dependent Fermi surface
that separates the filled and the empty regions in the phase
space. The fermions on the surface moving freely in the inverted
harmonic oscillator potential are described by, \be D_{t}v=\lambda
\,,~~~~D_{t}\lambda=v\,,\ee where $D_{t}$ denotes the co-moving
derivative following a phase point on the surface. Hence the free
fermions on the Fermi surface execute simple hyperbolic orbits.
However the time dependent profile of the Fermi surface is more
complicated. For small perturbations, the profile can be described
by the positions $v_{\pm}(\lambda,t)$ of the upper and lower
surfaces of the Fermi sea at each time, satisfying,\be
\partial_{t}v_{\pm}(\lambda,t)
=\lambda-v_{\pm}(\lambda,t)\partial_{\lambda}v_{\pm
}(\lambda,t)\,.\label{FermiSeaEOM} \ee Therefore, the static Fermi
sea satisfying the equation of motion (\ref{FermiSeaEOM}) is given
by, \be v_{\pm}(\lambda)=\pm
\sqrt{(\lambda^2-2\bar\mu)}\,,~~~i.e.~~~\frac{1}{2}(v^2-\lambda^2)+\bar
\mu=0\,,\ee which also given by the ground state of the
hamiltonian $H(v,\lambda) = (v^2-\lambda^2)/2+\bar\mu$ . For
simplicity, we assume that, in the ground state, the local maximum
of the potential is at zero energy level ($\mu_{c}=0$), and the
energy at the Fermi level is $\mu_F=-\mu$, measured from the local
maximum. Identifying \be v(\lambda)=\frac{\partial
\lambda}{\partial \tau}=\sqrt{(\lambda^2/\alpha'-2\bar\mu)}\ee as
the velocity of the classical trajectory of a particle at the
Fermi level, we define a new spatial coordinate $\tau$, the
classical time of motion at the Fermi level.

A single $D0$-brane state of the continuum theory corresponds to a
single eigenvalue or a fermion excited from the Fermi surface to
higher energy, {\it e.g.} to the top of the upside down potential.
This corresponds to putting the Neumann boundary condition on the
world sheet field $\tilde X(\sigma_{1}, \sigma_{2})$. Turning on
an exactly marginal boundary operator that interpolates between
the Neumann and Dirichlet boundary conditions on $\tilde
X(\sigma_{1}, \sigma_{2})$ makes the unstable brane to roll down
to D-objects localized at large Liouville direction $\phi$. This
corresponds to considering an unstable eigenvalue as the probe
fermion decaying in the presence of the $N$ free fermions, as it
executes classical Euclidean motion in the forbidden region along
the trajectory \be \lambda(t)=\sqrt{2\alpha'\bar\mu}~\hat \lambda
~\cos(t/\alpha')\,.\ee In order to relate to the picture
of the rolling tachyon in~\cite{sen-rolling}, the rolling
boundary state parametrized by $\tilde \lambda$, where $\sin\pi
\tilde\lambda=\hat\lambda$,  is identified with the decaying
eigenvalue starting at \be \lambda=-\sqrt{2 \alpha' \bar \mu}~\hat
\lambda \,, \ee where the energy of the state is given by \be
E=\bar\mu \cos^2 \pi \tilde \lambda\,. \ee  The time delay in the
classical evolution of the trajectory relative to the classical
trajectory at the Fermi level is therefore \be -\Delta
t=-\int^{\tau} \frac{d \tau'}{\sqrt{\alpha'}}=\int \frac{d \hat
\lambda}{\sqrt{\hat \lambda^2-1}}\simeq \ln\hat \lambda
\label{timedelay}\,,\ee for large $\hat \lambda$. The parameter
$\hat \lambda \in [0,1]$. The case $\hat \lambda=0$, {\it i.e.}
$\tilde \lambda=1$, describes an eigenvalue at the top of the
inverted harmonic oscillator potential ($V=0, E=-\bar \mu$) and
$\hat \lambda=1$, {\it i.e.} $\tilde \lambda=\frac{1}{2}$,
describes and eigenvalue on the Fermi sea ($V=-\bar \mu, E=0$) .

To relate such physical process in the free fermion theory to that
of the $D=1$ noncritical string theory in the dilaton and tachyon
background, one needs to bosonize the non-relativistic free
fermions in a suitable way. One way to view this is to express the
small fluctuations of the time dependent Fermi surface in terms of
the massless scalar $X(\tau , t)$, where $\tau =
\ln\hat\lambda$~(\ref{timedelay}) \cite{Polchinski-rev}. In this
respect equation (\ref{FermiSeaEOM}) can be generalized to the
time dependent case as

\be v_\pm (\lambda , t) = \mp\lambda \pm
\frac{1}{\lambda}\epsilon_\pm (\tau , t)\,, \ee

where \be \frac{1}{\sqrt\pi}\epsilon_\pm (\tau , t) = \pm
\partial_{t}X (\tau, t) - \partial_\tau X(\tau , t) \,. \ee The right
and left moving fluctuations $\epsilon_{\pm}(\tau,t)$ are
proportional to the right and left moving currents,
$J_{R}=:\psi_R^{\dagger}\psi_{R}:$ and
$J_{L}=:\psi_{L}^{\dagger}\psi_{L}:$ , generated by the chiral
fermionic variables as given by the bosonization relation \ba
&&\psi_R=\frac{1}{\sqrt{2 \pi}}:\exp \Big[i \sqrt{\pi}
\int^{\tau} d\tau' (\partial_{t}X-\partial_{\tau}X) \Big]: \nonumber\\
&&\psi_L=\frac{1}{\sqrt{2 \pi}}:\exp \Big[i \sqrt{\pi} \int^{\tau}
d\tau' (\partial_{t}X+\partial_{\tau}X) \Big]:  \ea (This is
because $(J_{L}+J_{R})=\sqrt{\pi} ~\partial_{\tau}X$ and
$(J_{L}-J_{R})=\sqrt{\pi}~ \partial_{t}X$) . The second quantized
hamiltonian for the free fermions is written in collective fields
~$\psi(\lambda, t)=\sum_{i} a_{i}\psi_{i}(\lambda) e^{-ie_{i}t}$
,~ where $\psi_i$ are the single particle wave function and
$a_{i}$ are the corresponding annihilation operators. One can
expand the fermionic field $\psi(\lambda, t)$ in $\psi_R$ and
$\psi_{L}$ as \be \psi(\lambda, t)=\frac{e^{i\bar \mu t}}{\sqrt{2
v(\lambda)}}\Big[\exp[{-i\int^{\lambda}d\lambda'
v(\lambda')+i\pi/4}]~\psi_L(\lambda,t)+\exp[{i \int^\lambda
d\lambda' v(\lambda')-i\pi/4}]~\psi_R(\lambda,t)\Big]\ee and
rewrite the second quantized hamiltonian \ba &&H \sim \int d\tau
\Big[i(\psi_R^\dagger \partial_\tau \psi_R-\psi_L^\dagger
\partial_\tau \psi_L)+\frac{1}{2 v^2}(\partial_\tau \psi_L^\dagger
\partial_\tau \psi_L+
\partial_\tau \psi_R^\dagger \partial_\tau \psi_R)
\nonumber\\
&&+\frac{1}{4}\big(\frac{v''}{v^3}-\frac{5 v'^2}{2
v^4}\big)(\psi_L^\dagger \psi_L+\psi_R^\dagger \psi_R)\Big]\,. \ea
Here $v'\equiv dv/d\tau$. Solving the equation (\ref{timedelay})
for $ \hat \lambda $ , $\hat
\lambda=\cosh(\tau/\sqrt{\alpha'})\simeq \exp{\tau},~\tau
\rightarrow \infty $ . Hence $v=\sqrt{2 \bar \mu} \sinh \tau
\simeq \sqrt {2 \bar \mu} \exp \tau$. This shows that the $
O(1/v^2)$ terms are not negligible in the double scaling limit
where we keep $\bar \mu $ fixed, unless we go to the asymptotic or
the $\tau $ $\rightarrow \infty$ region. This maps the second
quantized hamiltonian of $N$ non-relativistic free fermions to the
two dimensional Dirac hamiltonian. This shows that $\tau $ is the
natural spatial coordinate in terms of which the fermionic system
has a standard Dirac action to leading order in $\beta $ . Also in
terms of the bosonization variables the hamiltonian in the $\tau
\rightarrow \infty$ limit takes the form of the canonically
normalized free scalar hamiltonian \be H \sim \frac{1}{2}\int
d\tau \Big[ (\partial_t X)^2 + (\partial_\tau X)^2+
e^{-\tau}O(X^3)) \Big]\,, \ee with the equation of motion
$(\partial_t^2-\partial_\tau^2)X = e^{-\tau} O(X^2)$. This also
shows that $\tau$ can be identified with the zero mode of the
Liouville field~\cite{DasJevicki}. One can restrict $\lambda $ to
lie between the two turning points of the classical motion ( $ \pm
\sqrt{2 \alpha' \bar \mu}$ ) , {\it i.e.} $\hat \lambda \in
[-1,1]$ , or equivalently restricting $\tau $ to lie between $0$
and $T/2$ where $T$ is the period of classical motion ($\oint
d\tau=2\pi T$). One then imposes a standard bag-like boundary
conditions on the chiral fields to get a consistent picture.
Introducing the cut-off on $\tau $, the energy levels are
approximated by \be \epsilon_n \sim -\mu+n/T+\delta'(n/T,
\mu)+...\ee where the last two terms can be dropped in the limit
$T \rightarrow \infty $.

Thus, if a single fermion starts near the top of the upside down
potential it will roll down to infinity. In the region $\tau \gg 1
~, \hat \lambda \gg 1 $ , it becomes approximately relativistic
(since $v \gg 1 $). The final state could be expressed in terms of
the chiral fields bosonized as \be \psi_{L,R}(\tau,t)|0 \rangle =
e^{i~2 \sqrt{\pi} X_{L,R}(\tau,~t)}\,,\ee (where, $ X_L+X_R=X $ )
. Since the bosonization scalars have the mode expansion \be
X_{L,R}(\tau + t)=\int \frac{dp}{2 \pi \sqrt{2 E}} ~ \Big[
a_{L,R;p}^\dagger \exp \big(i |p|t+i p \tau \big)+\mbox{h.c.}
\Big]\,, \ee a final state of the decay would be given by \be \int
d \tau \Psi(\tau) \psi_L |0 \rangle= \int d \tau \Psi(\tau) e^{i~2
\sqrt{\pi} X_{L}(\tau,~t)} |0 \rangle=\int d \tau \Psi (\tau) e^{i
2 \sqrt{\pi}\int \frac {dp}{2\pi \sqrt{2 E}} e^{-i p \tau}}
a_{Lp}^\dagger |0 \rangle \,. \label{coherent state} \ee Thus
keeping in mind the form of the coherent state obtained in the
tree level string theory, $|\psi \rangle \sim \exp \big(\int \frac
{dp}{\sqrt{2E}} ~a_p^\dagger~ A~ |0\rangle \big)$, one can extract
the decay amplitude from the bosonization scalars as \be A \sim i
2 \sqrt {\pi} e^{-i E \tau }\,,~~|p|= E \,. \ee This is the same
as the amplitudes calculated from the continuum method.


\subsection{The RG Picture of Tachyonic Decay}

One can view such a process of decay of an unstable eigenvalue
describing the open string tachyon roll-down to $2D$ closed string
vacuum, as a Large $N$ Wilsonian RG in the singlet sector of the
matrix quantum mechanics leading to the $c=1$ fixed point (in a
similar fashion to our calculations in the previous sections). The
RG equations are nothing but a statement of response of the free
fermion system as successively one of the fermion is integrated
out in the presence of the rest of the free fermions. Hence,
following our calculation, by performing the Large $N$ world sheet
RG for the singlet Matrix quantum mechanics in the eigenvalue
representation (with
$\lambda_{N+1}(t)=z(t)=\sqrt{2\alpha'\bar\mu}~\hat \lambda~\cos
t/\sqrt{\alpha'}$), one can compare the change of the world sheet
free energy in the Large $N$ RG with that due to the closed string
emission picture. One can do this comparison both for the singlet
$MQM$ on infinite line and that on a large circle. In the later
case one would have to consider closed string emission amplitude
for compact time on large circle. This comparison, on one hand
gives rise to a physical picture of the Large $N$ RG we considered
in this paper, on the other hand it provides a way to test our
results. Note that the desired $c=1$ end point, corresponding to
the $2D$ closed string vacuum, always emerges in a very nice and
robust way in the RG calculation with all it's critical
properties.

We would now try to understand the emergence of this amplitude
from the point of view of the world sheet RG by turning on an
unstable eigenvalue \be z(t)= \sqrt {2\alpha'\bar \mu} ~\hat
\lambda ~\cos (t/\alpha')\,,\label{eiv}\ee and integrating it out
in the presence of the rest of the $N$ free fermions. The Fermi
sea, in response to this, would readjust its height
infinitesimally which is reflected in the renormalization of the
couplings. Later we will focus on the more general context of the
Wilsonian world sheet RG of the ~($N \times N$)~ arbitrary
matrices on circle rather than working in the eigenvalue
representation only, which is good for the large radius.

The partition functions before and after turning on this unstable
probe eigenvalue respectively, are given by,

\ba Z_{N}[g, t]=\int \prod_{i=1}^N d\lambda_i (t)
~\Delta_N(t)~\exp\Big[-\frac {N}{g^2} \int^{t} dt' [\frac{1}{2}
\dot \lambda_i(t')^2+V(\lambda_i(t'))]\Big]\,,
\nonumber\\
Z_{N+1}[g, t]=\int \prod_{i=1}^N d\lambda_i(t)
~\Delta_N(t)~\exp\Big[-\frac{(N+1)}{g^2}\int^{t} dt' [\frac {1}{2}
\dot \lambda_i(t')^2+V(\lambda_i(t'))]\Big]\nonumber\\
\int  dz(t)~ \exp\Big[-\frac{(N+1)}{g^2} \int^t dt'
\big\{\frac{1}{2} \dot z^2(t')+V(z(t'))\big\}\Big] ~\exp\big[~
\frac{1}{\hbar}~ \hat\phi(z(t))\big]\,, \label{Zamp1} \ea
 where the operator \be
\frac{1}{\hbar}~\hat\phi (z(t)) =
\log\Big[\frac{\Delta_{N+1}(t)}{\Delta_{N}(t)}\Big] =
\sum_{i=1}^{N} \log\Big[\lambda_i(t)-z(t)\Big]\,,~~~~N \sim
1/\hbar\,. \label{loop-op}\ee The Laplace transform of the
operator $\hat \phi (z(t))$ can be seen as the loop operator
$O(l)$ that creates a closed (Dirichlet) boundary of length $l$ on
the world sheet. Its expectation value gives the loop
amplitude~\cite{kazakov-kostov}. In the limit of the vanishing
boundary length, the operator becomes tachyonic and its
excitations correspond to the excitations of the closed string
tachyon: \ba \hat \phi(\tau,t)=\int_{-\infty}^{\infty} dq~
e^{-iqt}\int_0^{\infty} dl ~e^{-l \lambda(\tau)}~O(l,q)\,,\nonumber\\
O(l,q) \sim \frac{1}{2} \int dt~e^{iqt} \int d\tau ~e^{-l \lambda
(\tau)}~\partial_{\tau} X \,.\label{laplace}\ea The mode expansion
of the massless scalar field $X$ is related to the amplitude $A$
of the closed string decay in the sense of final state of the
evolution being somewhat like a coherent state \cite{kms}, as
shown in (\ref{coherent state}). Thus in the matrix model, one can
also compute the decay amplitude by computing $\hat \phi (\tau,
t)$. (The reference \cite{mgtv} describes a similar thought to
relate $\hat \phi(\tau, t)$ to the amplitude. However, they chose
$\hat \phi(\tau, t)$ to be in the same footing as $X$, instead of
(\ref{laplace})).

Let us now consider the partition function for the $N+1$
eigenvalues (\ref{Zamp1}) and rewrite the right hand side in terms
of the renormalized coupling $g'$ \ba Z_{N+1}[g, t] &=&  \int
\mathcal \prod_{i=1}^N d\lambda_i(t) ~
\Delta_N(t)~e^{-\frac{N}{g'^2}\int^{t} dt' [\frac {1}{2} \dot
\lambda_i(t')^2
+V(\lambda_i(t'))]}\nonumber\\
&&\int  dz(t)~ e^{-\frac{N}{g'^2} \int^t dt' [\frac{1}{2} \dot
z^2(t')+V(z(t'))]} ~e^{\frac{1}{\hbar}~\hat\phi(z(t))}\,, \ea
where $g' = g - \frac{g}{2N}+O(1/N^2)$. As before, extracting the
partition function of $N$ eigenvalues, we have \ba
\frac{Z_{N+1}[g, t]}{Z_{N}[g', t]} = \int  dz(t)~
\exp\Big[{-\frac{N}{g'^2} \int^t dt' \big\{\frac{1}{2} \dot
z^2(t')+V(z(t'))\big\}}\Big]
~\exp\big[~\frac{1}{\hbar}~\hat\phi(z(t))\big]
\nonumber\\
=\psi_t (N, g') \Big\langle \exp \big[~\frac{1}{\hbar}~\hat
\phi(z(t))\big] \Big\rangle\,, \label{Zamp2} \ea where $\psi_t (N,
g')$ is given by \be \psi_t (N, g')=\int dz(t)~
\exp\Big[-\frac{N}{g'^2} \int^t dt' \big\{\frac{1}{2} \dot
z^2(t')+V(z(t')\big\}\Big]\,. \ee Then from the RG point of view
the change of the world-sheet free energy ($\delta \mathcal F=
\log Z_{N+1}[g,t]/N^2-\log Z_N [g',t]/N^2$) in the process of
$D$-brane decay, given by the ratio \be \frac{Z_{N+1}[g,t]}{Z_ N
[g',t]}=\psi_t (N,g')~\big\langle \exp~\big[~\frac{1}{\hbar}~\hat
{\phi}(\tau,t)\big]\big \rangle \,, \label{amplitude} \ee encodes
the contribution from the amplitude $A$ through a knowledge of
$\hat \phi (\tau, t)$~. In terms of the discrete Callan-Symanzik
equation (\ref{C-S})~, the left hand side indicates a flow from $g
\rightarrow g'$ as $(N+1) \rightarrow N$. The right hand side
computes the contribution of the amplitude $A$ to the change of
the world sheet free energy as an O(1/N) effect. \be \Big[N \frac
{\partial}{\partial N}-\beta (g) \frac{\partial}{\partial
g}+\gamma \Big]\mathcal {F}[g,t] = \frac{1}{N}\langle \hat \phi
\rangle +\frac{1}{N^2}\ln \psi_t\,,~~~~\gamma=2\,.\label{CS-A} \ee
Here $\psi_t $ is the single particle wave function corresponding
to the probe eigenvalue. Note that for the singlet $MQM$ on a
large circle , the ~$\psi_t$ ~thus calculated by the Large $N$ RG
is a function of ($R+1/R$) which is a manifestation of the
anticipated $T$-duality property.

We would now  calculate the right hand side of the relation
(\ref{amplitude})~. Let us assume that the rest of the $N$
eigenvalues belong to the static Fermi sea forming a closed string
background given by \be \lambda_i=- \sqrt{2 \alpha' \bar
\mu}~~\hat \lambda\,,~~~~~i=1 \ldots N\,. \ee  Hence from
(\ref{loop-op}) and (\ref{eiv}) and considering $\hat \lambda \sim
\exp (\tau/\sqrt{\alpha'})$ for $\tau \gg 1 $ and also $N \sim
1/\hbar$ ,\be \hat \phi (\tau, t) = \ln\big[\sqrt{2 \alpha' \bar
\mu}(1-\cos \frac{t}{\sqrt{\alpha'}})\big]+\tau /\sqrt{\alpha'}\,.
\label{phi-hat}\ee Thus, \be \frac{Z_{N+1}[g,t]}{Z_N[g',t]}=\psi_t
(N,g') ~\big( \sqrt{2 \alpha' \bar \mu}(1-\cos
\frac{t}{\sqrt{\alpha'}})\big)^{N}~\Big \langle \exp~
\tau/\hbar~\sqrt{\alpha'}~\Big \rangle \,. \ee Here the $t$
dependent part could be absorbed in the overall prefactor of the
ratio. As in (\ref{amplitude}) effectively $\hat \phi \sim
\tau/\sqrt{\alpha'}$ contains the contribution from the decay
amplitude to the change of the world sheet free energy.

We will now go back to the relation of $\hat\phi$ with the
bosonization field $X$ (and hence $A$) and verify that the
expression of $\hat \phi$ ~we obtained in (\ref{phi-hat}) does
lead to the standard answer for $A$. Using (\ref{laplace}) and
performing the delta function integration over $t$~, \ba \hat \phi
(\tau,t) = \pi \int d\tau' \frac{1}{\sqrt{2 \alpha' \bar
\mu}~(\exp{\tau/\sqrt{\alpha'}}
+\exp{\tau'/\sqrt{\alpha'}})}~(\partial_{\tau'}X (\tau,t))\,.~\ea
This implies~, \be (\partial_{\tau'}\hat \phi) = \frac{\pi
(\partial_{\tau'} X)}{\sqrt{2 \alpha' \bar
\mu}~(\exp{\tau/\sqrt{\alpha'}}+\exp{\tau'/\sqrt{\alpha'}})}\,.\ee
Inverting and solving for $X$, \be X=\frac{1}{\pi}\int d \tau'
\sqrt{2 \alpha' \bar
\mu}~(\exp{\tau/\sqrt{\alpha'}}+\exp{\tau'/\sqrt{\alpha'}})
(\partial_{\tau'} \hat \phi)\,. \ee Considering $(\partial_{\tau'}
\hat \phi) = \frac{1}{\sqrt{\alpha'}}~\delta (\tau'-\tau)$, we
have~$X= \frac{2}{\pi} ~\sqrt{2\bar\mu} ~e^{
\frac{\tau}{\sqrt{\alpha'}}}$~. Comparing with the relation of the
mode expansion of $X$ to the amplitude $A$ we have, \be A \sim
\frac{2}{\pi} ~\sqrt{2\bar\mu} ~e^{ \frac{\tau}{\sqrt{\alpha'}}}
\sim e^{\ln \hat \lambda}\,. \ee The time delay
(\ref{timedelay})is thus given by $|\Delta t| \sim \ln \hat
\lambda$ which is consistent. Thus we conclude that in our RG,
integration of one eigenvalue in the presence of the others
describes the decay of an unstable $D0$-brane with open string
tachyon attached to it to the $2D$ closed string theory (with its
$D0$-branes) with an amplitude $A$ given by  \be \Big[N \frac
{\partial}{\partial N}-\beta (g) \frac{\partial}{\partial
g}+\gamma \Big]\mathcal {F}[g,t] = \frac{1}{N}\ln A
+\frac{1}{N^2}\ln \psi_t\,,~~~A \sim
e^{\ln\hat\lambda}\,,~~~\gamma=2\,. \ee In our set up it is clear
that there is a flow, $g \rightarrow g'$ as $(N+1) \rightarrow N$,
corresponding to the decay process. However, as we have discussed
in the previous chapter, the eigenvalue representation is too
simple to explicitly produce a produce a nontrivial fixed point of
the flow ($\beta (g)=-g/2$) from the beta function equation. The
flow in the eigenvalue representation only gives the gaussian
fixed point corresponding to the inverted harmonic oscillator.
However, still one can say that the flow is hitting the $c=1$
fixed point situated at the infinitesimal distance from the
Gaussian fixed point. Working in a more general set up of $MQM$ of
general $N \times N$ matrices, one can explicitly show from the
beta function that the endpoint of the flow is indeed a pair of
$c=1$ nontrivial fixed point ($\beta(g)=-g/2+3F(g,R)~g^3/2$)~dual
to the $2D$ closed string theory the $D$-brane decay leads to.
Also it is shown that for $R \to \infty$ or even for large finite
$R$ (where the singlet free fermion picture is meaningful) the
pair of $c=1$ fixed points are drawn infinitesimally close to each
other and to the Gaussian fixed point such that they are
overlapping. Thus it is consistent to think that even in the
simple eigenvalue representation the flow corresponding to the
$D$-brane decay does end at the pair of overlapping $c=1$ fixed
point in the infinitesimally close neighborhood of the gaussian
fixed point. This is derived in detail in the previous sections
and is the key observation of this paper.

We will now discuss the decay process in the context
of RG analysis of the general
$N \times N$ matrices $\phi_N(t)$. Consider the usual
partition function $Z_{N+1}$ of the matrix field $\phi_{N+1}(t)$
that is decomposed into $\phi_N(t)$ and the $(1\times N)$ and
$(N\times 1)$ row and column vectors $v^*(t)$ and $v(t) $ respectively,

\ba
&&Z_{N+1}[g,M]=\int \mathcal D^{N^2} \phi_N(t)~ \mathcal D^N
v^*(t) ~\mathcal D^N v(t) \nonumber\\ &&\exp
\Big[-N~\mbox{Tr}\int dt~ \big[ \frac{1}{2}(D
\phi_N)^2+ \frac{1}{2}M^2\phi_N^2 -\frac{g}{3} \phi_N^3 + D v^*
D v+ M^2 v^* v -g~ v^* \phi_N v \Big]\,. \label{boundary}
\nonumber\\
\ea
As usual, the covariant derivative is defined in terms of the
non-dynamical gauge field $A$ in the open string spectrum
corresponding to the vertex operator $\dot t$, \be D
\phi_N=\partial_t \phi_N - [ A, \phi_N ]\,. \ee The gauge field
$A$ projects onto the $SU(N)$ singlet wave functions by acting as
a Lagrange multiplier.

Now, one can see $Z_{N+1}$ as the dual large $N$ description of
the $2D$ string theory in the presence of the space-time filling
$D1$-branes, the extended Liouville boundary states tensored with
Neumann boundary state for the (Euclidean) time. One can similarly
understand $Z_{N+1}$ for the (Euclidean) time taken on a (large)
circle. Motivated from previous works
\cite{yang-open,minahan-open,kazakov-kostov}, the authors of
\cite{kms} pointed out that integration over the vectors
$v^*(t),~v(t)$ inserts into the random surfaces dynamical
boundaries (quark loops) wandering in the time direction. Although
their model is essentially the same as (\ref{boundary}), in the
former case the couplings in the flavour part are introduced by
hand rather than being determined by the original closed string
action due to the parametrization (\ref{matrixpara}).

As we have discussed in
previous sections, precisely this is happening in our RG. The
$(N+1) \times (N+1)$-th matrix element gives the position of the
wandering boundaries, which for simplicity, have been set to zero
in our analysis . The sum of the {\it `single line'} one loop
Feynman diagrams, generated by the integration over the vectors
$(v^*, v)$~ (see equations~(\ref{SigmaFourier})~and~
(\ref{SigmaInvFourier}))~, inserts in the path integral different
operators coupled to the dynamical boundaries.
The relevant operators coupled to these boundaries are
of the form $\frac{1}{N} (\mbox{Tr} \phi^n-g_n v^* \phi^{(n-2)} v )$.
The massless open
string tachyon would correspond to the operators $\int dt~
e^{iqt}~ v^*v(t)$. For a model without the kinetic term for
$\big(v^*, v\big)$, integration over the vectors generates the
macroscopic loops, the boundaries with the Dirichlet boundary
condition on time. This happens when both the parameters
$M^2$ and $g$ acquire large values due to tuning. As
pointed out  in~\cite{kms} , tuning the parameters $M^2$ and $g$
together with $N$ one should be able to get a scaling model with
two independent parameters $\mu_B$ and $\mu$, the renormalized
boundary and bulk cosmological constants, given by the relation
(\ref{uniformization}).
In our RG analysis we arrive at an
analogous relation (\ref {Mscaling}) with usual scaling between
$\mu_B$ and $\mu$ (with an accuracy of 80 \%). This presumably
indicate the presence of various boundaries in $2D$
string theory. However to realize whether these boundaries
correspond to Neumann or Dirichlet case needs further investigation.
Here, tuning of
the matrix coupling constant ~$g$~ gives rise to a renormalized
bulk cosmological constant $\mu$. The scaling variable
$\Delta=(1-g/g^*)$ depends nontrivially on $\mu$

\be
\Delta \sim
\frac{1}{2\pi} \mu \ln \mu \,,
\ee
as they are mutually coupled
through the Fermi-Dirac distribution of the fermions in the grand
canonical ensemble ~\cite{GK1}.
On the other hand one can think of
tuning the mass parameter $M^2$ to control the boundary cosmological
constant $\mu_B$ as

\be
\mu_B \sim1-M^2/M^{*2}\,.
\ee
Hence solving the equation

\be
\frac{\beta_g}{\beta_M}=\frac{\partial g}{\partial M}=
f(g,M,T)\,,
\ee
the RG flows $ g=g(M,T)$ in the Double scaling
limit near the nontrivial fixed points is expected to give rise to
relations analogous to the trajectory
(\ref{uniformization}). In equation (\ref{Mscaling}) we have
solved this trajectory for large $R$ (for matrix quantum mechanics
on circle) and got a consistent scaling. We expect the fixed
points of large $M^*$ obeying such scaling could be Dirichlet
boundaries. However, since in our  analysis we are dealing with
bosons on a circle and do have kinetic term for the quarks, these
could be Neumann boundaries as well. We hope to come back to these
interesting questions in close future.

\section{Discussion}
\setcounter{equation}{0}

In our analysis we have seen the following remarkable fact.
The general framework of world-sheet
RG analysis of the matrix quantum mechanics
on a circle is seen to be capable
of reproducing the critical exponent and
the expected scaling laws of the $c=1$ fixed point
and the $T$-duality property respected by
its singlet sector. Moreover, it indirectly reflects the presence
of the non-singlet sector from the tendency of the change
of the world-sheet free energy to change sign at a critical
radius. It also indicates the presence of the boundaries
in the $2D$ string theory and gives a flavor of capturing
their dynamics. All these serve to be an initial understanding
and progress towards our main goal, to develop a scheme
for dealing with the non-singlet sectors (or the world-sheet
vortices) which govern the physics of $2D$ string theory on
small circle. In a separate publication \cite{DD-bh} we will describe
how the behavior of the non-singlet sector can be explicitly captured
in our RG scheme with a gauged matrix model, by introducing
an appropriate gauge breaking term and tuning the fugacity of the
vortices  coupled to it (along with the usual tuning of the other
parameters in the double scaling limit). In this case, the RG flows
will turn out to be capable to explore the interesting
fixed points beyond the $c=1$ universality class.

As we have observed,  by tuning the parameters $M^2$ and $g$
we arrive at RG trajectories with appropriate scaling of the bulk
and the boundary cosmological constants, suggesting the presence
of the various boundaries of the underlying $2D$ theory.
To explicitly see these boundaries for each of the Dirichlet and
Neumann cases one needs to study the behavior of the operator creating
boundaries of finite length. The most relevant operator that controls this
length is the boundary cosmological constant.  Such finite length
boundary operators are computed from the free fermion wave function.
Using our RG scheme one can study the behavior of these operators
for the $c=1$ fixed point at any radius and try to see the all possible
boundaries.
The results at large $R$ can be compared with the free fermion wave 
functions 
as a test. The study can be further extended to include the effects
of the non-singlets. We have work in progress in this area.

We would now comment on the subtle issues on dealing
the nonlocal terms in the action, in computing the determinant
emerging from the adiabatic integration over the vector row and column
of the original matrix variable. In our method, we are expanding the
determinant in the Fourier space about $\phi(t)=0$ (small field
approximation).
We tackle the non-local
integrals by introducing 'center of mass' (large) and 'relative' (small)
coordinates,
expanding all the functions around the latter and then
integrating over the 'relative' coordinates. We
keep the contribution from the non-localities in the action up to the
kinetic term $\Pi^2(t)$, and the contribution
from the higher order terms in $\phi(t)$ up to the $\phi^3(t)$ term.
The kinetic term is the most important nonlocal term to capture the
effect of the world sheet vortices. As by the parametrization
(\ref{matrixpara})
it also gives rise to the kinetic term of the vector fields, it
plays an important role in
understanding the  nonlocal Neumann boundaries, the $D1$ brane.
The higher order terms like $\mbox{Tr}(\Pi^m(t) \phi^n(t))$
generated by expanding the determinant give rise to powers of traces
and are redundant for the RG analysis. The higher order terms are also
negligible due to small field approximation. However, it will be interesting
to
study the scheme with more couplings to see the contribution from the higher
order terms and the convergence of the scheme. Also it is an interesting
question
to understand the relevant operators of the flow from the matrix quantum
mechanics
point of view.

An interesting solution of the two-dimensional theory, apart from the
flat space background with linear dilaton, is the two-dimensional
black hole \cite{rabi,wadia-bh,witten-cigar}. It is partially
believed that the nonperturbative formulation of two-dimensional
string theory in terms of an integrable theory of noninteracting
nonrelativistic fermions of the matrix quantum mechanics can deal
issues like black hole evaporation and gravitational collapse. In
any case, the condensation of world-sheet vortices of the
two-dimensional noncritical string theory should describe
two-dimensional black hole background, which had been a long
standing challenge to obtain from matrix model. The recent work by
Kazakov, Kostov and Kutasov gives proposals for a matrix model for
two-dimensional black hole \cite{KKK}. In our forthcoming paper
\cite{DD-bh}, we will address this issue in more detail.
Also in this context, one can consider integrating out several
vector rows and columns simultaneously to see the effect of
inserting multiple boundaries. This is analogous to the situation
of the decay of several unstable branes and is anticipated by some
authors to give black-hole like phase at the end point of the
decay \cite{akk,GutperleKraus}. However, the multiple row and
column integration presumably would insert redundant operators of the form
$(\mbox{Tr}\phi_N ^m)^n$, which are not likely to
give rise to black hole fixed point.

\newpage

{\bf Acknowledgments}

We would like to thank Michael Douglas for useful discussions, comments
and constant support at all stages of the work. We would also like to thank
Abhishek Agarwal, Igor Klebanov, Massimo Porrati and especially Edouard
Br\'ezin for illuminating discussions and useful comments
on a preliminary draft. The research of SD was supported in part by
DOE grant DE-FG02-96ER40959. The research of TD was supported
in part by NSF grants PHY-0070787 and PHY-0245068. Any opinions, findings, and
conclusions or recommendations expressed in this material are those of the
authors and do not necessarily reflect the views of the National Science
Foundation.

\addcontentsline{toc}{section}{Appendix}
\appendix{The Feynman Diagrams}\label{A1}

Here we evaluate and discuss the terms in different orders of the
series $\Sigma [g,M,\phi_N,R,N]$ (\ref{SigmaFourier}) using the
summation rules discussed in section (3.3) and the relation
(\ref{FourierTrfm}) for the inverse Fourier Transform.

\subsection {The terms of order O($\phi$)}

\be
g~\mbox{Tr}\bigg[\sum_n\frac{1}{(\frac{n^2}{R^2}+M^2)}~\phi_0\bigg]=
\frac{g}{2M}\coth (\pi M R) \int_{0}^{2 \pi R}dt
~\mbox{Tr}~\phi(t)\,.
 \ee

\subsection {The terms of order O($\phi$$\phi$)}

\ba &&\frac{g^2}{2}\mbox{Tr}~\bigg[\sum_{m,n}\frac{1}
{\big(\frac{m^2}{R^2}+M^2
\big)\big(\frac{n^2}{R^2}+M^2\big)}~\phi_{m-n}\phi_{n-m}\bigg]
\nonumber\\
&&=\frac{g^2}{2} \int \frac{dt_{1} dt_{2}}{(2 \pi R)^2}
~\mbox{Tr}~( \phi(t_{1})\phi(t_{2}))
\bigg[\sum_{m,n}\frac{\exp\big(i(n-m)~t_{1}/R\big)
\exp\big(i(m-n)~t_{2}/R\big) }{\big(\frac{m^2}{R^2}+M^2
\big)\big(\frac{n^2}{R^2}+M^2\big)} \bigg]\,.
\nonumber\\
\ea Now, changing the variables to 'center of mass' and 'relative'
coordinates defined respectively by \be
T=\frac{t_{1}+t_{2}}{2}\,,~~~~~\tau=\frac{t_{1}-t_{2}}{2}\,, \ee
we have, \be dt_{1}~dt_{2}=    J  \Big(\frac{t_{1},t_{2}}{T,
\tau}\Big)~dT~d\tau=2~dT~d\tau\,. \ee Hence, \ba
&&\frac{g^2}{2}~\mbox{Tr}\bigg[\sum_{m,n}\frac{1}{\big(\frac{m^2}{R^2}+M^2
\big)\big(\frac{n^2}{R^2}+M^2\big)}~\phi_{m-n}\phi_{n-m}\bigg]
\nonumber\\
&&=\frac{g^2}{4 \pi^2 R^2} \int_{0}^{2 \pi R} dT \int_{-\pi
R}^{\pi R}d\tau ~\mbox{Tr}( \phi\big(T+\tau)\phi(T-\tau)\big)
\sum_{m,n}\frac{\exp\big(i(n-m)~2\tau/R\big)}{\big(\frac{m^2}{R^2}+1
\big) \big(\frac{n^2}{R^2}+1\big)}
\nonumber\\
&&\simeq \frac{g^2}{4 \pi^2 R^2} \int_{0}^{2 \pi R} dT \int_{-\pi
R}^{\pi R}d\tau ~\mbox{Tr}( \phi(T)^2-\tau^2 \dot \phi(T)^2)
\sum_{m,n}\frac{\exp\big(i(n-m)~2\tau/R\big)}{\big(\frac{m^2}{R^2}+1
\big)\big(\frac{n^2}{R^2}+1\big)}
\nonumber\\
&& \simeq g^2 F_{g2} (R, M) \int_{0}^{2 \pi R} dT~\mbox{Tr}
(\phi(T)^2/2)+g^2 \hat F_{g2} (R, M) \int_{0}^{2 \pi R} dT
~\mbox{Tr}(\dot \phi(T)^2/2)\,, \ea where, \ba &&F_{g2} (R,
M)=\frac{1}{2 \pi^2 R^2}\int_{-\pi R}^{\pi R} d\tau \sum_{m,n}
\frac{ \exp\big(i(m-n)~2\tau/R\big)}{\big(\frac{m^2}{R^2}+M^2
\big) \big(\frac{n^2}{R^2}+M^2\big)}
\nonumber\\
&&=\frac{1}{2 \pi^2 R^2}\int_{0}^{2 \pi R} d\tau' \sum_{m,n}
\frac{\exp\big(i(m-n)~\tau'/R\big)}{\big(\frac{m^2}{R^2}+M^2 \big)
\big(\frac{n^2}{R^2}+M^2\big)}\,,~~~~(\tau'=2 \tau)
\nonumber\\
&&=\frac{1}{2 M^2 \sinh^2 \pi M R} \int _{0}^{2 \pi R} d\tau'
\cosh M (\pi R+  \tau') \cosh M (\pi R- \tau')
\nonumber\\
&&=\frac{1}{M^3 \sinh^2 \pi M R} \bigg[\frac{1}{2} \pi M R \cosh
2\pi  M R+\frac{1}{8}\sinh 4\pi  M R\bigg]\,, \ea and, \ba &&\hat
F_{g2}(R, M)=-\frac{1}{2 \pi^2 R^2}\int_{-\pi R}^{\pi R}
d\tau~\tau^2 \sum_{m,n} \frac{
\exp\big(i(m-n)~2\tau/R\big)}{\big(\frac{m^2}{R^2}+M^2 \big)
\big(\frac{n^2}{R^2}+M^2\big)}
\nonumber\\
&&=-\frac{1}{8 \pi^2 R^2}\int_{0}^{2 \pi R} d\tau'~\tau'^2
\sum_{m,n}\frac{
\exp\big(i(m-n)~\tau'/R\big)}{\big(\frac{m^2}{R^2}+M^2 \big)
\big(\frac{n^2}{R^2}+M^2\big)} \,,~~~~(\tau'=2 \tau)
\nonumber\\
&&=-\frac{1}{8 M^2 \sinh^2 \pi M R}\int_{0}^{2 \pi R}
d\tau'~\tau'^2 \cosh M (\pi R+\tau') \cosh M (\pi R-\tau')
\nonumber\\
&& =\frac{1}{M^5 \sinh^2 \pi M R}\bigg[-\frac{1}{64} (1+8 \pi^2
M^2 R^2) \sinh 4\pi M R -\frac{\pi^3 M^3 R^3}{6}\cosh 2\pi M R
\nonumber\\
&&~~~~~~~~~~~~~~~~~~~~~~~+\frac{\pi M R}{16} \cosh 4 \pi M R
\bigg]\,.
\nonumber\\
\ea

\subsection {The terms of order O($\phi$$\phi$$\phi$)}

\ba
&&\frac{g^2}{6}\mbox{Tr}~\bigg[\sum_{m,n,k}\frac{1}{\big(\frac{m^2}{R^2}+M^2
\big)\big(\frac{n^2}{R^2}+M^2\big)\big(\frac{k^2}{R^2}+M^2\big)}
~\phi_{m-n}\phi_{n-k}\phi_{k-m}\bigg]
\nonumber\\
&&=\frac{g^2}{6} \int \frac{dt_{1} dt_{2} dt_{3}}{(2 \pi R)^3}
~\mbox{Tr}~( \phi(t_{1})\phi(t_{2})\phi(t_{3}))
\nonumber\\
&&\bigg[\sum_{m,n}\frac{\exp\big(i(n-m)~t_{1}/R\big)
\exp\big(i(k-n)~t_{2}/R\big)
\exp\big(i(m-k)~t_{1}/R\big)}{\big(\frac{m^2}{R^2}+M^2
\big)\big(\frac{n^2}{R^2}+M^2\big)\big(\frac{k^2}{R^2}+M^2\big)}
\bigg]\,.
\nonumber\\
\ea

Using redefinition of the variables into the "center of mass" and
the "relative coordinates",
$$
T=\frac{1}{3}(t_{1}+t_{2}+t_{3})\,,
~~~\tau_{1}=(t_{1}-t_{2})\,,~~~~\tau_{2}=(t_{1}-t_{3})\,,
$$
$$
 dt_{1} ~dt_{2} ~dt_{3}
 = J \Big(\frac{t_{1},t_{2},t_{3}}{T,\tau_{1},\tau_{2}}
 \Big)~dT~d\tau_{1} ~d\tau_{2}=dT ~d\tau_{1} ~d\tau_{2}\,.
$$
Considering $\tau_{1}$ and $\tau_{2}$ to be small and keeping the
order $O( \phi^3)$ term, above series could be evaluated as, \ba
&&\frac{g^3}{6}~\mbox{Tr}~\bigg[\sum_{m,n,k}\frac{1}{\big(\frac{m^2}{R^2}+M^
2
\big)\big(\frac{n^2}{R^2}
+M^2\big)\big(\frac{k^2}{R^2}+M^2\big)}~\phi_{m-n}\phi_{n-k}\phi_{k-m}\bigg]
\nonumber\\
&&=\frac{g^3}{48 \pi^3 R^3} \int _{0}^{2 \pi R} dT~ \int_{- \pi
R}^{\pi R} d\tau_{1} ~\int_{- \pi R}^{\pi R} d\tau_{2}
\nonumber \\
&&\mbox{Tr}\Big( \phi(T+\frac{\tau_{1}+\tau_{2}}{3})
\phi(T-\frac{2}{3}\tau_{1}+\frac{1}{3}
\tau_{2})\phi(T+\frac{1}{3}\tau_{1}-\frac{2}{3}\tau_{2})\Big)
\nonumber\\
&&\Bigg[\sum_{m,n,k}\frac{\exp \big(i n \tau_{1} /R\big) \exp
\big(-i m \tau_{2}/R\big) \exp \big(i k (\tau_{2}-\tau_{1})/R
\big)} {\big(\frac{m^2}{R^2}+M^2\big)\big(\frac{n^2}{R^2}+M^2\big)
\big(\frac{k^2}{R^2}+M^2\big)} \Bigg]
\nonumber\\
&& \simeq g^3 F_{g3}(R,M) \int _{0}^{2 \pi R} dT~\mbox{Tr}
(\phi(T)^3/3)  \ea where, \ba && F_{g3}(g,R)
\nonumber\\
&&=\frac{1}{16 \pi^3 R^3}\int_{- \pi R}^{\pi R} d\tau_{1} ~\int_{-
\pi R}^{\pi R} d\tau_{2} \Bigg[\sum_{m,n,k}\frac{ \exp \big(i n
\tau_{1} /R\big) \exp \big(-i m \tau_{2}/R\big) \exp \big(i k
(\tau_{2}-\tau_{1})/R \big)}
{\big(\frac{m^2}{R^2}+M^2\big)\big(\frac{n^2}{R^2}+M^2\big)
\big(\frac{k^2}{R^2}+M^2\big)}  \Bigg]
\nonumber\\
&&=\frac{1}{16 M^2 \pi R \sinh^2\pi M R}\int_{- \pi R}^{\pi R}
d\tau_{1} ~\int_{- \pi R}^{\pi R} d\tau_{2} ~\cosh(\pi M R -M
\tau_1)~\cosh(\pi M R+ M \tau_2)
\nonumber \\
&&~\sum_k \frac{\exp \big(i k (\tau_{2}-\tau_{1})/R \big) }
{\big(\frac{k^2}{R^2}+M^2\big)}
\nonumber\\
&&=\frac{1}{64 M^2 \pi R \sinh^2\pi M R} \int_{- 2 \pi R}^{2 \pi
R} d\hat T ~\int_{-2 \pi R}^{2 \pi R} d\hat \tau ~\big(\cosh(2 \pi
M R +M \hat \tau)+\cosh M \hat \tau \big)~\sum_k \frac{\exp i k
\hat \tau /R}{\big(\frac{k^2}{R^2}+M^2\big)}
\nonumber\\
&&=\frac{\pi R}{16 M^3 \sinh^3\pi M R } \int_{0}^{2 \pi R} d\hat
\tau \Big[ \big(\cosh(2 \pi M R +M \hat \tau)+\cosh M \hat \tau
\big)~\cosh(\pi M R-M \hat \tau)
\nonumber\\
&&+\big(\cosh(2 \pi M R -M \hat \tau)+\cosh M \hat \tau
\big)~\cosh(\pi M R+M \hat \tau)\Big]
\nonumber\\
&&=\frac{\pi M R}{64 M^5 \sinh^3\pi M R (\cosh 2 \pi M R + \cosh 4
\pi M R ) }~~[ 4 \pi M R \big(3 \cosh \pi M R + 2 \cosh3 \pi M R
\nonumber \\
&&+2 \cosh 5 \pi M R + \cosh 7 \pi M R \big) + \sinh \pi M R +
\sinh 3 \pi M R+ \sinh 5 \pi M R+ 2 \sinh 7 \pi M R
\nonumber \\
&&+ \sinh 9 \pi M R ] \ea

\appendix{The Scaling Dimensions}\label{A2}

Here we will summarize the general expession of the matrix
$\Omega_{k,l}=\frac{\partial \beta_k (\Lambda^*)}{\partial \Lambda_l}$
and its eigenvalues $\Omega_1$ and $\Omega_2$,
the scaling dimensions of the relevant operators, at different fixed points.

\ba
&&\Omega_{11}=\frac{1}{2}(5h-1)+3 g^{*2} (F_{g3}^*-\frac{3}{2} \hat
F_{g2}^*)
\nonumber\\
&&\Omega_{12}=g^{*3}\Big(\frac{\partial F_{g3}^*}{\partial M}-\frac{3}{2}
\frac{\partial \hat F_{g2}^*}{\partial M}\Big)
\nonumber\\
&& \Omega_{21}=g^*\Big((1/M^*-M^*) F_{g2}^*-M^* \hat F_{g2}^*\Big)
\nonumber\\
&&\Omega_{22}=h+g^{*2} \frac{\partial}{\partial M}\Big((1/2M^*-M^*/2)
F_{g2}^*-M^*/2 \hat F_{g2}^*\Big)
\ea

For the gaussian fixed point
\be
\Omega_{11}=(5h-1)/2=-1/2  \,,~~~~\Omega_{22}=h=0
\,,~~~~\Omega_{12}=\Omega_{21}=0\,.
\ee

For the nontrivial fixed point
\be
\Omega_{11}=1+5h/2=1 \,,~~~~\Omega_{22}\simeq h\simeq
0\,,~~~~\Omega_{12}\simeq \Omega_{21}\simeq 0\,.
\ee


\end{document}